\documentclass[bibyear]{aa}
\usepackage{graphicx}

\usepackage{natbib}
\bibpunct{(}{)}{;}{a}{}{,} 

\usepackage{txfonts}
\usepackage{longtable}
\usepackage{listings}
\usepackage{color}
\usepackage{textcomp}

\begin{document}

\title{Clusters and mirages:\\ cataloguing stellar aggregates in the Milky Way \thanks{Full Table~1 and the list of individual members are available in electronic form only
at the CDS via anonymous ftp to cdsarc.u-strasbg.fr (130.79.128.5)
or via http://cdsweb.u-strasbg.fr/cgi-bin/qcat?J/A+A/} }

   \author{T. Cantat-Gaudin\inst{\ref{UB}} and
F. Anders\inst{\ref{UB}}
        }

  \institute{
                Institut de Ci\`encies del Cosmos, Universitat de Barcelona (IEEC-UB), Mart\'i i Franqu\`es 1, E-08028 Barcelona, Spain\label{UB}
                \\
                 \email{tcantat@fqa.ub.edu}
                 }

   \date{Received October 12, 2018; accepted ???, ???}

  \abstract{Many of the open clusters listed in modern catalogues were initially reported by visual astronomers as apparent overdensities of bright stars. As observational techniques and analysis methods continue to improve, some of them have been shown to be chance alignments of stars and not true clusters. Recent publications making use of \textit{Gaia}~DR2 data have provided membership lists for over a thousand clusters, however, many nearby objects listed in the literature have so far evaded detection.}{We aim to update the \textit{Gaia}~DR2 cluster census by performing membership determinations for known clusters that had been missed by previous studies and for recently-discovered clusters. We investigate a sub-set of non-detected clusters that, according to their literature parameters, should be easily visible in \textit{Gaia}. Confirming or disproving the existence of old, inner-disc, high-altitude clusters is especially important as their survival or disruption is linked to the dynamical processes that drive the evolution of the Milky Way.}{We employed the \textit{Gaia}~DR2 catalogue and a membership assignment procedure, as well as visual inspections of spatial, proper motion, and parallax distributions. We used membership lists provided by other authors when available.}{We derived membership lists for 150 objects, including ten that were already known prior to \textit{Gaia}. We compiled a final list of members for 1481 clusters. Among the objects that we are still unable to identify with the \textit{Gaia} data, we argue that many (mostly putatively old, relatively nearby, high-altitude objects) are not true clusters.}{At present, the only confirmed cluster located further than 500\,pc away from the Galactic plane within the Solar circle is NGC~6791. It is likely that the objects discussed in this study only represent a fraction of the non-physical groupings erroneously listed in the catalogues as genuine open clusters and that those lists need further cleaning.}

\titlerunning{Cataloguing stellar aggregates in the Milky Way}
\authorrunning{T. Cantat-Gaudin and F. Anders}
\maketitle

\section{Introduction}

Our ancestors have been gazing at the night sky since ancient times, observing the stars, identifying and memorising their patterns and cycles. Despite the scientific revolutions, paradigm shifts, and instrumental breakthroughs of the past millennia, astronomy has a long history of continuity in its terminology and conceptual tools. For instance, the modern, official division of the celestial sphere into 88 constellation adopted by the International Astronomical Union in 1922 is mostly based (at least in the Northern hemisphere) on Ptolemy's \textit{Almagest}\footnote{The work of \citet{1984ptal.book.....T,1998ptal.book.....T} is considered its most faithful and scrupulous English translation.} (written circa 150~AD), one of the most influential scientific publications of all time, which remained a reference throughout the Middle Ages \citep{2012A&A...544A..31V}. The \textit{Almagest} itself both relied on and superseded the work of previous astronomers, such as Eudoxus of Cnidus, who introduced Greece to concepts of Babylonian astronomy he had studied in Egypt (for instance, the division of the ecliptic into twelve zodiac constellations).

Stellar clusters are among the most obvious celestial objects. Some are visible to the naked eye; and archaeological findings, such as the Lascaux cave \citep[circa 17,000~BC,][]{1997ascu.conf..217R,2001EM&P...85..391R} or the Nebra disc \citep[circa 1600~BC,][]{2003JRASC..97..245M}, suggest that ancient populations had observed the open cluster now known as the Pleiades. Nearly half of the currently-known open clusters were catalogued by Charles Messier \citep{1781cote.rept..227M}, William Herschel \citep{1786RSPT...76..457H,1789RSPT...79..212H,1802RSPT...92..477H} and John Herschel \citep{1864RSPT..154....1H}, and they were included in the New General Catalogue\footnote{ \citet{2010ocns.book.....S} points out that the ``Other Observers'' column in Dreyer's original NGC paper refers to 180 discoverers and early observers, including Caroline Herschel, Nicolas-Louis de Lacaille, Amerigo Vespucci, Hipparchus of Nicaea, and Aratus of Soli.} \citep[NGC,][]{1888MmRAS..49....1D}. A few more clusters were identified when the use of photographic techniques became common in the late 19th century and several authors added their own discoveries to lists of previously reported objects \citep[e.g.][]{1895MmRAS..51..185D,1910MmRAS..59..105D,1908AnHar..60..199B,1915MmRAS..60..175M,1930LicOB..14..154T,1931AnLun...2....1C}. Numerous objects have been discovered since then and subsequently added to catalogues of open clusters \citep[e.g.][]{1958csca.book.....A,1970csca.book.....A,1982A&A...109..213L,1985IAUS..106..143L,1995ASSL..203..127M,2002A&A...389..871D,2013A&A...558A..53K,2019AJ....157...12B}.

As groups of coeval stars, open clusters are useful laboratories for the study of stellar evolution \citep[e.g.][]{1983ApJS...51...29V,2007ApJ...669.1167B,2013aspm.confE...9S,2017MNRAS.466.2161B,2018ApJ...863L..33M}. 
They have been used as convenient probes of the structure and evolution of the Galactic disc \citep[e.g.][]{1930LicOB..14..154T,1973A&A....23..317M,1982ApJS...49..425J,1995ARA&A..33..381F,2010IAUS..266..106M,2016EAS....80...73M} and its metallicity gradient \citep[e.g.][]{1979ApJS...39..135J,1997AJ....114.2556T,2012AJ....144...95Y,2009A&A...494...95M,2016A&A...588A.120C,2016A&A...591A..37J,2017MNRAS.470.4363C, 2018AJ....156..142D}.

Photometric and astrometric studies of young clusters reveal clues about stellar formation processes \citep{2018MNRAS.474.1176J,2019ApJ...870...32K}, while old objects hold fossil information about the past of our Galaxy \citep[e.g.][]{1994AJ....107.1079P,2006AJ....131.1544B,2008A&A...488..943S,2015AJ....150..200H}.

Astrometric datasets containing proper motions (and sometimes parallaxes) allow us to identify clusters as overdensities in higher-dimensional spaces than just their projected 2D distribution on the sky. Examples of such studies include: 
\citet{1999A&A...345..471R} \citep[using Hipparcos data,][]{1997A&A...323L..49P}; 
\citet{2003A&A...410..565A} \citep[with Tycho-2 data,][]{2000A&A...355L..27H};
 \citet{2012A&A...543A.156K} \citep[using PPMXL proper motions,][]{2010AJ....139.2440R}; 
\citet{2017MNRAS.470.3937S} \citep[using UCAC4 proper motions,][]{2013AJ....145...44Z};
\citet{2018A&A...615A..49C} \citep[from \textit{Gaia}~DR1,][]{2016A&A...595A...2G}.

The second data release \citep[DR2,][]{2018A&A...616A...1G} of the ESA \textit{Gaia} space mission \citep[][]{2016A&A...595A...1G} is by far the deepest and most precise astrometric catalogue ever obtained, with proper motion nominal uncertainties a hundred times smaller than UCAC4 and PPMXL. In a systematic search for known clusters in the \textit{Gaia}~DR2 catalogue, \citet{2018A&A...618A..93C} were only able to identify 1169 objects, a surprisingly low number given that more than 2000 optically visible clusters are listed in the literature \citep{2002A&A...389..871D,2013A&A...558A..53K} and given that \citet{2017MNRAS.470.3937S} reported potential members for 1876 objects based on UCAC4 proper motions alone. Further investigation of the literature available for these objects revealed that the existence of many of them had already been questioned \citep[notably by][when building the Revised New General Catalogue]{1973rncn.book.....S} or even convincingly refuted \citep[e.g. four NGC objects by][]{2018MNRAS.480.5242K}.

This paper investigates clusters for which no membership list is available from the \textit{Gaia}~DR2 data in Sect.~\ref{sec:membership}. 
Section~\ref{sec:asterisms} focuses on some of the non-recovered objects that, according to their literature parameters, should actually be easily detected in the \textit{Gaia}~DR2. We argue that these objects are asterisms rather than physical clusters. 
Section~\ref{sec:bias} contains considerations on the propagation of non-verified objects in the literature. Section~\ref{sec:discussion} discusses the consequence of the non-existence of these objects for the Galactic census and our understanding of the Milky Way. Finally,  Sect.~\ref{sec:conclusion} presents our concluding remarks.

\section{Membership determinations} \label{sec:membership}

\subsection{Data and method} 

The first step in our search for known clusters that had been missed by \citet{2018A&A...618A..93C} was to cross-match the list of members proposed by \citet{2017MNRAS.470.3937S} for these objects with the \textit{Gaia}~DR2 data. In most cases, the proper-motion distribution of the putative members form a coherent group within the nominal uncertainties of UCAC4 ($\sim$5\,mas\,yr$^{-1}$ at $G$=14 and $\sim$10\,mas\,yr$^{-1}$ at $G$=16) but it is very scattered in the \textit{Gaia}~DR2 data (which features proper motion uncertainties of $\sim$3$\times10^{-2}$\,mas\,yr$^{-1}$ at $G$=14). For ten of them, however, visual inspection revealed a clump of co-moving stars in proper motion space. These objects (listed in Table~\ref{tab:meanparams}) were further analysed with the UPMASK membership determination procedure  \citep[][]{2014A&A...561A..57K}. 

Since the procedure verifies the compactness of the groups in positional space, using an inappropriately small field of view results in undetected clusters. \citet{2018A&A...618A..93C} relied on the apparent sizes quoted by \citet{2002A&A...389..871D} (hereafter DAML) and \citet{2013A&A...558A..53K} (hereafter MWSC) to perform cone searches to the \textit{Gaia} archive.
 \citet{2014A&A...561A..57K} show that a field of view corresponding to 1 to 2 times the size of the cluster (defined as the distance at which it becomes indistinguishable from the field) is a reasonable choice.
The present study managed to recover several objects by significantly increasing the radius of the investigated field of view (e.g. 24' radius for NGC~2126, where DAML quotes a total radius of 6', or 40' for Collinder~421 instead of 3.6').

We queried the \textit{Gaia}~DR2 data through the ESAC portal\footnote{https://gea.esac.esa.int/archive/}, and scripted most queries using the package \texttt{pygacs}\footnote{https://github.com/Johannes-Sahlmann/pygacs}. Following the procedure of \citet{2018A&A...618A..93C}, we did not apply any quality filtering \citep[such as the filters proposed by][]{2018A&A...616A..17A}, but we only queried the stars brighter than $G$=18 with a 5-parameter astrometric solution. This magnitude cut roughly selects the $\sim$20\% sources with the most precise astrometry.

The UPMASK code was originally developed for photometric classification and it has been successfully applied to astrometric data \citep[e.g.][]{2018A&A...615A..49C,2018A&A...618A..93C}. Its principle relies on grouping stars according to their parallax and proper motion (in this implementation we use k-means clustering) and, in a second step, verifying whether the distribution of these stars on the sky is more concentrated than what can be expected from random fluctuations in a uniform distribution (in this implementation, we use the total length of a minimum spanning tree). The procedure is repeated multiple times and at each iteration, the proper motions and parallaxes used for the clustering are randomly sampled from the probability distribution function of the astrometric parameters of each star (the 3D normal distribution corresponding to the nominal uncertainties \texttt{pmra\_error}, \texttt{pmdec\_error}, \texttt{parallax\_error}, and the correlation coefficients listed in the \textit{Gaia}~DR2 catalogue). Stars that are classified as a member of a concentrated group at most iterations are attributed a higher clustering score that can be interpreted as a membership probability.

\begin{figure*}[ht!]
\begin{center} \includegraphics[scale=0.8]{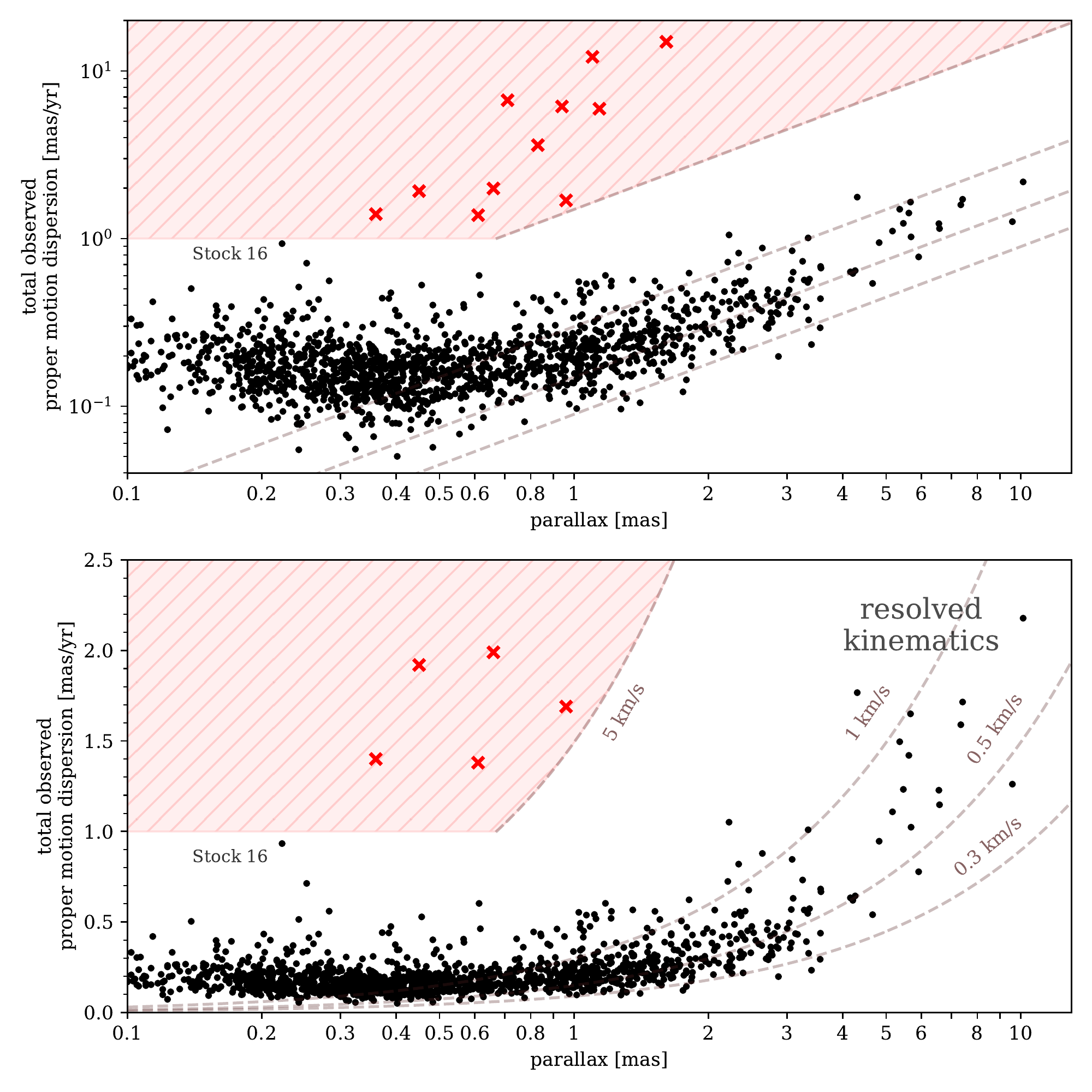} \caption{\label{fig:totPMdisp_vs_par}  Top: total proper-motion dispersion against mean parallax for clusters identified in the \textit{Gaia}~DR2 data (blue dots) and the asterisms for which membership lists are available (red crosses). The dashed lines show the theoretical proper-motion dispersion corresponding to 1D velocity dispersions of 0.3, 0.5, 1, and 5\,km\,s$^{-1}$ in the absence of any measurement error. The shaded area indicates the region of the parameter space where we consider a group cannot be a physical cluster (see Sect.~\ref{sec:asterisms}). Bottom: same as top, but with a decimal vertical scale.} \end{center}
\end{figure*}

\subsection{Adding members of recently discovered clusters} \label{sec:found}

Many of the clusters that were recently discovered with \textit{Gaia} data are not listed in \citet{2018A&A...618A..93C}. We collected the membership determinations provided by the authors for the UBC clusters \citep[University of Barcelona,][]{2018A&A...618A..59C,2019A&A...627A..35C}, the UFMG clusters \citep[Universidade Federal de Minas Gerais,][]{2019MNRAS.483.5508F}, and the COIN clusters \citep[Cosmostatistics Initiative][]{2019A&A...624A.126C} and performed determinations for the clusters whose membership was not made available by the respective authors. We do not include the 76 candidate clusters reported by \citet{2019arXiv191012600L}, which were published as we applied the final revisions to this paper.

We used UPMASK to determine members for the  Gaia~1 and Gaia~2 clusters \citep[discovered with \textit{Gaia}~DR1 by][]{2017MNRAS.470.2702K} for which the authors do not list membership probabilities. We did not add the two objects (Dias~4a and Dias~4b) reported by \citet{2018MNRAS.481.3887D} which turn out to match the coordinates, distance, and age of NGC~5269 and SAI~118, as quoted by DAML and MWSC, and which are not, therefore, new clusters.

A more recent study by \citet{2019JKAS...52..145S} identified 207 objects within 1\,kpc. The authors do not provide a membership list, so we applied UPMASK to these objects as well. Although all 207 do correspond to clear overdensities in astrometric space, many of them only correspond to weakly defined spatial concentrations and we were only able to compute membership probabilities for 141 of them. Three of those turned out to have been reported before: UPK~19 (UBC~32), UPK~176 (UBC~10a), and UPK~327 (UBC~88). We provide membership probabilities for the remaining 138.

Many of the UPK objects reported by \citet{2019JKAS...52..145S} are spatially very sparse and are reminiscent of the large-scale structures identified by \citet{2019AJ....158..122K} or several of the groups identified in proper motion space by \citet{2019A&A...624A.126C}. We show the spatial distribution, colour-magnitude diagram, and proper motions of six selected UPK clusters (illustrating their variety in density, age, and morphology) in Figs.~\ref{fig:upk_85} to \ref{fig:upk_649}.

Table~\ref{tab:meanparams} summarises the mean astrometric parameters of the ten clusters known prior to \textit{Gaia}~DR2. The electronic version of this Table contains:

a) the ten known clusters whose membership is established in this study;
b) 138 UPK clusters reported by \citet{2019JKAS...52..145S}, with membership probabilities computed in this study;
c) Gaia~1 and Gaia~2 from \citet{2017MNRAS.470.2702K}, whose membership is established in this study;
d) 1225 of the 1229 clusters whose membership was published by \citet{2018A&A...618A..59C}. The excluded entries are: BH~140 and FSR~1758 (that the paper showed to be globular clusters), FSR~1716  \citep[another globular,][]{2017ApJ...838L..14M,2017A&A...605A.128K} that is erroneously included in the study as it is not flagged as such in MWSC, and Harvard~5 (a duplicate of Collinder~258);
e) 46 clusters (including 41 COIN-Gaia clusters) whose members were published in \citet{2019A&A...624A.126C};
f) 57 UBC clusters whose members were published in \citet{2018A&A...618A..59C} and \citet{2019A&A...627A..35C};
g) three UFMG clusters whose members were published by \citet{2019MNRAS.483.5508F};
for a total of 1481 objects. 

As an electronic table, we also provide the list of individual members (436,242 stars with non-zero membership probability) for each of these 1481 clusters.

\begin{table*}
\begin{center}
        \caption{ \label{tab:meanparams} Summary of mean parameters for the OCs that have been newly characterised in this study. Full table of 1481 clusters confirmed by {\it Gaia} DR2, as well as the table of individual cluster member candidates, are available as an electronic table via the CDS.}
        \small\addtolength{\tabcolsep}{-1pt}
        \begin{tabular}{ c  c  c  c  c  c  c  c  c  c  c  c  c  c }
        \hline
        \hline

OC & $\ell$ & $b$ & $\alpha$ & $\delta$ & r$_{50}$ & $N$ & $\mu_{\alpha}*$ & $\sigma_{\mu_{\alpha}*}$ & $\mu_{\delta}$ & $\sigma_{\mu_{\delta}}$ & $\varpi$ & $\sigma_{\varpi}$ & $d$\\
  & [deg] & [deg] & [deg] & [deg] & [deg] &   & [mas\,yr$^{-1}$] & [mas\,yr$^{-1}$] & [mas\,yr$^{-1}$] & [mas\,yr$^{-1}$] & [mas] & [mas] & [pc]\\
        \hline
BH~205 & 344.632 & 1.632 & 254.053 & -40.636 & 0.097 & 96 & -0.15 & 0.145 & -1.083 & 0.193 & 0.569 & 0.065 & 1672  \\
Berkeley~100 & 113.657 & 2.459 & 351.485 & 63.781 & 0.022 & 39 & -3.372 & 0.186 & -1.557 & 0.181 & 0.123 & 0.07 & 6579  \\
Collinder~421 & 79.453 & 2.523 & 305.829 & 41.701 & 0.143 & 167 & -3.651 & 0.123 & -8.334 & 0.113 & 0.813 & 0.048 & 1187  \\
FSR~0451 & 115.748 & -1.121 & 357.955 & 60.916 & 0.214 & 231 & -3.216 & 0.107 & -1.907 & 0.083 & 0.32 & 0.049 & 2862  \\
Harvard~20 & 56.312 & -4.686 & 298.321 & 18.345 & 0.079 & 46 & -1.732 & 0.089 & -4.447 & 0.084 & 0.461 & 0.050 & 2040  \\
NGC~2126 & 163.23 & 13.15 & 90.658 & 49.883 & 0.100 & 119 & 0.848 & 0.112 & -2.615 & 0.103 & 0.747 & 0.043 & 1287  \\
NGC~2169 & 195.631 & -2.92 & 92.125 & 13.951 & 0.076 & 65 & -1.068 & 0.187 & -1.655 & 0.171 & 0.982 & 0.083 & 989  \\
NGC~2479 & 235.998 & 5.359 & 118.762 & -17.732 & 0.075 & 129 & -4.318 & 0.100 & 1.053 & 0.078 & 0.626 & 0.058 & 1527  \\
Ruprecht~65 & 263.077 & -1.533 & 129.838 & -44.041 & 0.09 & 40 & -4.746 & 0.081 & 4.245 & 0.058 & 0.412 & 0.038 & 2268  \\
Ruprecht~8 & 226.153 & -3.901 & 105.424 & -13.539 & 0.147 & 63 & -1.02 & 0.099 & -1.424 & 0.087 & 0.444 & 0.047 & 2115  \\
        \hline
        \hline
        \end{tabular}
\tablefoot{$N$: number of stars with membership probabilities over 50\%. $d$: mode of the distance likelihood after adding a parallax offset of +0.029\,mas.}
\end{center}
\end{table*}

\subsection{Intrinsic and apparent proper-motion dispersion} \label{sec:intrinsic}

Although the apparent proper-motion dispersion of a cluster does not constitute an accurate diagnostic of its dynamical state, we argue that it can be a sufficient empirical basis to discriminate between plausible and implausible clusters.  

The internal velocity dispersion of a bound stellar system depends on its mass and physical size. Dispersions in the core of globular clusters typically reach 5 to 10\,km\,s$^{-1}$ \citep[e.g.][]{1993ASPC...50..357P,2015ApJ...798...23L,2018MNRAS.478.1520B}. As they are less massive systems, open clusters are expected to exhibit smaller dispersions. Line-of-sight velocities obtained from high-resolution spectroscopy shows that Trumpler~20 \citep{2014A&A...561A..94D}, NGC~6705 \citep{2014A&A...569A..17C}, M~67 \citep{2016BaltA..25..432V}, Trumpler~23 \citep{2017A&A...598A..68O}, or Pismis~18 \citep{2019A&A...626A..90H} have internal 1D velocity dispersions below 2\,km\,s$^{-1}$ (possibly much less given the measurement uncertainty on individual velocities). For an object at a distance of 1\,kpc, this upper limit corresponds to a proper-motion dispersion of $\sim$0.4\,mas\,yr$^{-1}$.

All the clusters mentioned in Sect.~\ref{sec:found} can be identified in the \textit{Gaia}~DR2 astrometric data and, in particular, they all exhibit a compact proper-motion distribution. Their total proper-motion dispersion (quadratic sum of the dispersion in \texttt{pmra} and \texttt{pmdec}) is shown in Fig.~\ref{fig:totPMdisp_vs_par} as a function of mean parallax, along with the theoretical proper-motion dispersion for intrinsic velocity dispersions of 0.3, 0.5, and 1\,km\,s$^{-1}$.

For clusters more distant than $\sim$500\,pc (parallax smaller than 2\,mas), the uncertainty on the \textit{Gaia}~DR2 proper motions of the members starts contributing significantly to the observed dispersion and clusters more distant than 1\,kpc are dominated by measurement uncertainties. The uncertainty on the membership status and contamination by field stars also artificially increase the observed proper motion dispersion, especially for the most distant objects. 

\begin{figure*}[ht!]
\begin{center} \resizebox{0.9\textwidth}{!}{\includegraphics[scale=0.6]{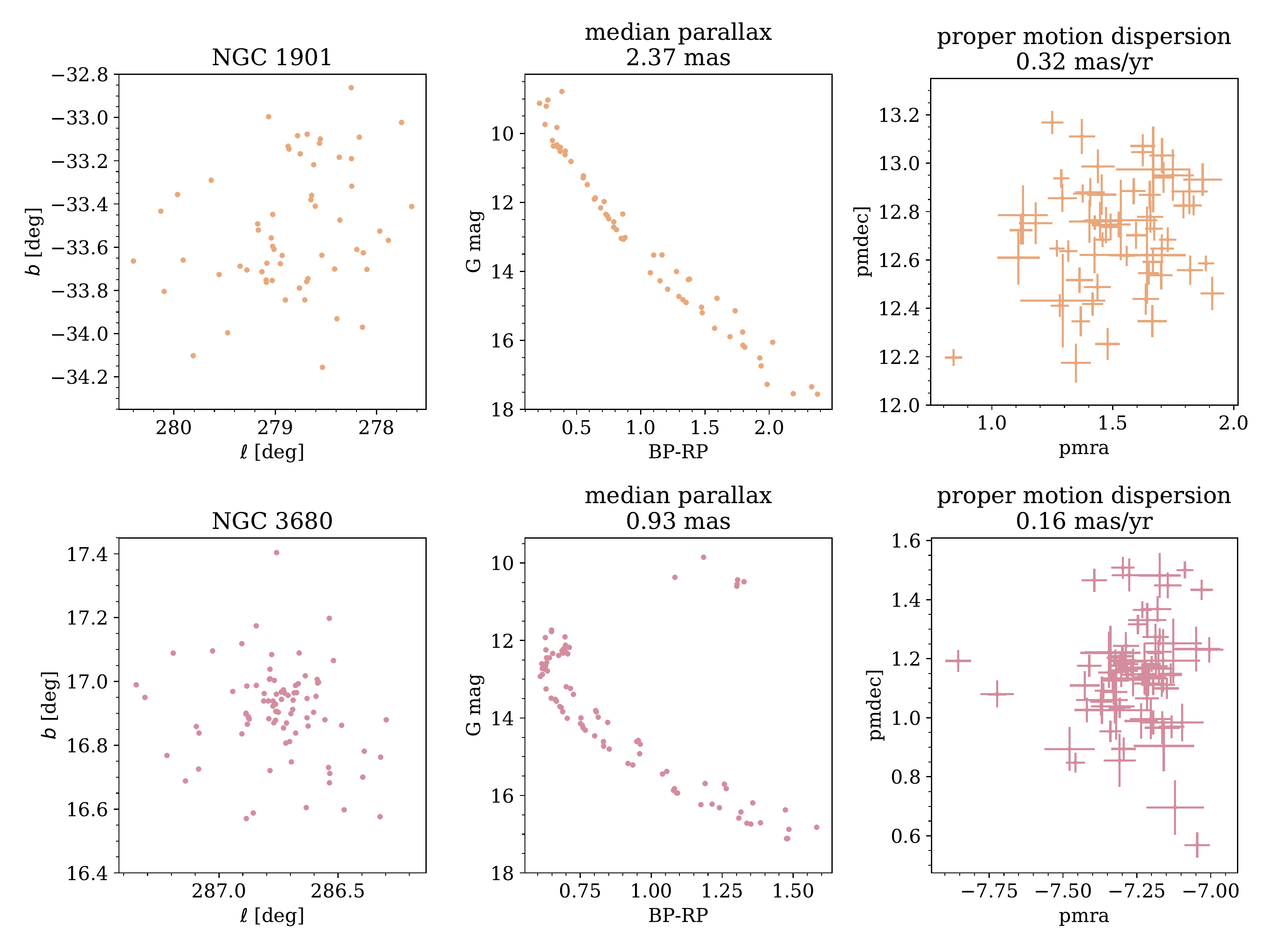}} \caption{\label{fig:tworemnants} Probable members (probability > 70\%) of the archetypal cluster remnants NGC~1901 and NGC~3680. Left: sky position. Middle: colour-magnitude diagram. Right: proper motions.} \end{center}
\end{figure*}

Since the total mass of clusters decreases as they age (because of escaping stars and stellar evolution) and the stars with the highest velocities are ejected first, their velocity dispersion also decreases \citep{2001MNRAS.321..199P}. In \textit{Gaia}~DR2, the sparse, high-altitude, outer disc objects NGC~1901 and NGC~3680 \citep[considered archetypes of dynamically evolved clusters by][]{2001A&A...366..827B} exhibit proper-motion dispersions of 0.3 and of 0.2\,mas\,yr$^{-1}$, respectively (see Fig.~\ref{fig:tworemnants}). The radial velocities of \citet{1997A&A...322..460N} also show a line-of-sight velocity dispersion under 1\,km\,s$^{-1}$ for NGC~3680. A more nearby example of a dynamically evolved cluster is Ruprecht~147 \citep[2.5\,Gyr old at a distance of $\sim$310\,pc according to][]{2018A&A...619A.176B}, which \citet{2019AJ....157..115Y} estimate to have lost as much as 99\% of its initial mass. The proper-motion dispersion we observe for this cluster is 0.65\,mas\,yr$^{-1}$, which at this distance translates into a 1D velocity dispersion smaller than 1\,km\,s$^{-1}$. The seven evolved van den Bergh-Hagen clusters studied by \citet{2016MNRAS.463.3476P} all exhibit small proper-motion dispersions as well. 

In Fig.~\ref{fig:totPMdisp_vs_par}, the distant grouping with the largest proper-motion dispersion is Stock~16, a very young embedded aggregate projected against the tip of a molecular pillar in the HII region RCW~75 \citep{1977A&AS...30..307F,1985ApJ...292..148T,2005A&A...430..471V,2014MNRAS.442.3761N}. Stock~16 appears substructured and is likely part of a larger complex of young stars. Its proper-motion dispersion of nearly 1\,mas\,yr$^{-1}$ should therefore be considered close to an upper limit for what can be realistically expected from a bound stellar system.

\section{Non-physical groupings} \label{sec:asterisms}

The recent compilation of clusters and candidates by \citet{2019AJ....157...12B} is the most up-to-date compiled catalogue of clusters\footnote{This catalogue is however only complete up to the clusters identified in \textit{Gaia}~DR1 data by \citet{2017MNRAS.470.2702K} and \citet{2018A&A...618A..59C} and contains none of the subsequent UBC, Gulliver, UFMG, and UPK clusters.} and contains 10,978 entries, only 1644 of which are flagged as asterisms. This list contains a large number of cluster candidates identified in infrared surveys (invisible to \textit{Gaia}, as discussed in Sect.~\ref{sec:discussion}). None of the 4968 entries flagged as `embedded' have been detected so far in the \textit{Gaia} data.

However, \citet{2019AJ....157...12B} also list many objects that should easily be visible to \textit{Gaia} (in particular, all the NGC objects since they were discovered by direct observation at the eyepiece) but have so far remained undetected.
In this Section, we focus on 38 objects (listed in Table~\ref{table:asterisms}) that are expected to be relatively nearby clusters (1 to 2\,kpc), most of them at high Galactic latitudes. Since they should be easily seen but have so far remained undetected in the \textit{Gaia} data, we argue that they are not true clusters. Only six of them are flagged as asterisms in \citet{2019AJ....157...12B}, and four of those six were shown to be non-existent by \citet{2018MNRAS.480.5242K} beyond a reasonable doubt on the basis of \textit{Gaia}~DR2 astrometry and ground-based spectroscopic observations. Table~\ref{table:asterisms} attempts to provide an exhaustive list of studies mentioning these 38 asterisms.

\subsection{Trying to identify elusive and neglected clusters}

Many of the NGC clusters listed in Table~\ref{table:asterisms} are flagged as non-existent in the Revised New General Catalogue \citep{1973rncn.book.....S} or were questioned by various authors. Some are absent from the widely-used DAML and MWSC catalogues and some from the WEBDA database \citep{1995ASSL..203..127M}. The reality of some of these objects has been the subject of debate and controversy (e.g. NGC~1252 or NGC~6994), but others are simply mentioned as `neglected' or `poorly-studied' by authors who might have not been aware that others have expressed doubts about their existence.

The lack of clear spatial concentrations of stars near the coordinates reported in the catalogues has led several authors to posit that some of these objects might be the remnants of dynamically evolved, dissolved open clusters. The age estimates available in the literature for these objects are often over 1\,Gyr, making this claim plausible. The uncertainty on the proper motions available before \textit{Gaia}~DR2 was not sufficient to identify co-moving groups beyond a few hundred parsecs, and the reality of these clusters was mainly argued on the basis of subjective patterns in colour-magnitude diagrams (CMDs). The idea that some of the objects flagged as remnants might not be clusters at all was already put forward by \citet{2006BASI...34..153C} (for NGC~6994), \citet{2010A&A...510A..44M} (for NGC~6863) and \citet{2018MNRAS.480.5242K} (for NGC~1252, NGC~6994, NGC~7772, and NGC~7826).

The objects discussed here are part of a larger list of clusters that we were unable to find in the \textit{Gaia}~DR2 data\footnote{In total, over 150 objects not flagged as asterisms or embedded clusters in \citet{2019AJ....157...12B} have not yet been detected with \textit{Gaia}~DR2 data.}. For this study, we focused on these 38 objects in particular because their existence has been questioned in the past, because  their elusiveness has been justified as their being cluster remnants, or because catalogues lists them at distances under 3\,kpc and high Galactic latitudes, which should make their detection easy in the \textit{Gaia} data. 

We provide detailed comments on why we consider these 38 objects to be asterisms in Sects.~\ref{indiv:start} to \ref{indiv:end}.

\subsection{Discarding groups from their proper-motion dispersion}

Seven of the objects that we argue are really asterisms were recently investigated by authors who provide lists of members. The \textit{Gaia}~DR2 proper motions and parallaxes of those proposed members are shown in Appendix~\ref{sec:indiv_asterisms}.

For the purpose of this study we establish a simple but quantitative criterion based on proper-motion dispersions. A given set of sources might potentially be a physical cluster if it fulfills any of these two conditions: 1) Its proper-motion dispersion corresponds to a physical velocity dispersion of less than 5\,km\,s$^{-1}$. This value is very permissive because such high dispersions are only observed in globular clusters and it can realistically only be expected for systems hosting many thousands of solar masses (as opposed to a questionable grouping of a handful of stars);
2) Its observed proper-motion dispersion is less than 1\,mas\,yr$^{-1}$. This value is about three times the contribution of the \textit{Gaia}~DR2 measurement errors and this ensures we do not discard objects whose apparent proper-motion dispersion is dominated by these errors.

We show in Fig.~\ref{fig:totPMdisp_vs_par} that the proposed lists of members for NGC~1663, NGC~2180, NGC~3231, NGC~6481, NGC~7036, NGC~7193, and Ruprecht~3 do not fulfill any of the two  conditions described above.

\begin{table*}
\begin{scriptsize}
\begin{center}
        \caption{ \label{table:asterisms} Table of asterisms. }
        \begin{tabular}{ c c c c c c c c c }
        \hline
        \hline
Name & $\ell$ & $b$ & d$_{\mathrm{DAML}}$ & d$_{\mathrm{MWSC}}$ & WEBDA &  \citet{2019AJ....157...12B} & considered & considered\\
     & [deg] & [deg] & [pc] & [pc] & & {\bf} & real by & dubious by\\
    \hline   
    \hline     
NGC~1252 & 274.08 & -50.83 & 790 & 944 & yes & asterism & {\citet{1983PASP...95..474B}} & {\citet{1973rncn.book.....S}}\\
 & & & & & & & {\citet{2001A&A...366..827B}}             & {\citet{1984PASP...96...70E}}                 \\
 & & & & & & & {\citet{2001A&A...374..554P}}             & {\citet{1998A&A...340..402B}}                 \\
 & & & & & & & {\citet{2003ARep...47....6L}}             & {\citet{2018MNRAS.480.5242K}}                 \\
 & & & & & & & {\citet{2005ApJ...619..824X}}             & {\citet{2019A&A...624A...8A}}                 \\
 & & & & & & & {\citet{2007A&A...468..139P}}             &               \\
 & & & & & & & {\citet{2011MNRAS.412.1611P}}             &               \\
 & & & & & & & {\citet{2012A&A...548A..97Z}}             &               \\
 & & & & & & & {\citet{2013MNRAS.434..194D}}             &               \\
    \hline 
NGC~1520 & 291.14 & -35.70 & 775 & 1023 & no & OC &     &  {\citet{1973rncn.book.....S}}  \\
    \hline    
NGC~1557 & 283.77 & -38.26 & 1055 & 1820 & no & OC & {\citet{2001A&A...366..827B}} & {\citet{1973rncn.book.....S}}   \\
 & & & & & & & {\citet{2011JKAS...44....1T}} &     \\
    \hline    
NGC~1641 & 277.20 & -38.32 & 985 & 985 & yes & OC & {\citet{2001A&A...366..827B}} &  {\citet{1963IrAJ....6...74S}}  \\
 & & & & & & & {\citet{2006JKAS...39..115K}} &     \\
    \hline   
NGC~1663 & 185.85 & -19.74 & 700 & 1490 & yes & OC & {\citet{2003A&A...407..527B}} & {\citet{2010A&A...516A...3K}} \\
 & & & & & & &                                   {\citet{2007A&A...468..139P}} &     \\
 & & & & & & &                                   {\citet{2019A&A...624A...8A}} &     \\
    \hline    
NGC~1746 & 179.07 & -10.65 & 800 & 800 & yes & OC & {\citet{1937AnHar.105..403C}}    & {\citet{1992BaltA...1..125S}}\\
 & & & & & & &     & {\citet{1998A&A...333..471G}} \\
 & & & & & & &     & {\citet{1998A&AS..131...89T}} \\
 & & & & & & &     & {\citet{2010PASP..122.1008L}} \\
 & & & & & & &     & {\citet{2018A&A...618A..93C}} \\
    \hline    
NGC~1963 & 240.99 & -30.87 &    & 1703 & no & OC &  {\citet{2001A&A...366..827B}}   &  {\citet{1973rncn.book.....S}}  \\
         &         &        &      &      &    & &     & {\citet{2011JKAS...44....1T}}\\
    \hline    
NGC~2132 & 268.70 & -30.18 & 974 & 1003 & no & OC &     &  {\citet{1973rncn.book.....S}}  \\
    \hline    
NGC~2180 & 203.91 & -7.10 & 910 & 1882 & yes & OC & {\citet{2004A&A...427..485B}} & {\citet{1973rncn.book.....S}}\\
 & & & & & & & {\citet{2008A&A...477..165P}} &     \\
 & & & & & & & {\citet{2011MNRAS.412.1611P}} &     \\
 & & & & & & & {\citet{2019A&A...624A...8A}} &     \\
    \hline    
NGC~2220 & 252.50 & -23.93 & 1170 & 1393 & no & OC &  & {\citet{1973rncn.book.....S}}   \\
    \hline    
NGC~2348 & 278.14 & -23.81 & 1070 & 1076 & no & OC & {\citet{2001A&A...366..827B}} & {\citet{1973rncn.book.....S}}   \\
    \hline     
NGC~2394 & 210.78 & 11.47 & 940 &    & yes & asterism & {\citet{2006JKAS...39..115K}} &  {\citet{1973rncn.book.....S}}  \\
    \hline      
NGC~3231 & 141.95 & 44.60 & 715 &    & no & OC & {\citet{2019A&A...624A...8A}} & {\citet{1973rncn.book.....S}}\\
 & & & & & & & \citet{2011JKAS...44....1T} &  \\
  & & & & & & & {\citet{2012A&A...542A..68P}} &  \\
    \hline    
NGC~4230 & 298.03 & 7.45 & 1445 & 2630 & yes* & OC & {\citet{2011JKAS...44....1T}} &    \\
 & & & & & & & \citet{2019MNRAS.488.4648P} &     \\
    \hline  
NGC~5269 & 308.96 & -0.67 & 1410 & 1634 & no & OC & {\citet{2017MNRAS.466.4960P}}    & {\citet{1973rncn.book.....S}}\\
         &         &        &      &      &    & &     & {\citet{2011JKAS...44....1T}}\\
    \hline    
NGC~5998 & 343.80 & 19.83 & 1170 & 4853 & no & OC & {\citet{2011JKAS...44....1T}} & {\citet{1973rncn.book.....S}}   \\
    \hline    
NGC~6169 & 339.39 & 2.52 & 1007 & 1007 & yes* & OC &     & {\citet{1973A&AS...10..135M}}\\
 & & & & & & &     & {\citet{2011JKAS...44....1T}} \\
    \hline    
NGC~6481 & 29.94 & 14.94 & 1180 &    & no & OC & {\citet{2007A&A...468..139P}} &  {\citet{1973rncn.book.....S}}  \\
 & & & & & & & {\citet{2019A&A...624A...8A}} &     \\
    \hline    
    
NGC~6525 & 37.4 & 15.91 & 1436 & 3221 & no & OC & \citet{2019MNRAS.488.4648P} & {\citet{1973rncn.book.....S}}   \\
         &         &        &      &      & &    &     & {\citet{2010A&A...516A...3K}}\\
         &         &        &      &      & &    &     & {\citet{2011JKAS...44....1T}}\\

    \hline  
NGC~6554 & 11.67 & 0.65 & 1775 & 1775 & no & OC &       & {\citet{1973rncn.book.....S}}\\
         &         &        &      &      & &    &     & {\citet{2011JKAS...44....1T}}\\
    \hline    
NGC~6588 & 330.84 & -20.88 & 2314 & 4757 & no & OC & {\citet{2011JKAS...44....1T}} & {\citet{1973rncn.book.....S}}   \\
 & & & & & & & {\citet{2015NewA...38...31C}} &     \\
 & & & & & & & {\citet{2017NewA...51...15M}} &     \\
    \hline    
NGC~6573 & 9.05 & -2.09 & 460 & 3032 & no & OC & {\citet{2018MNRAS.477.3600A}} &  {\citet{1973rncn.book.....S}}  \\
    \hline    
NGC~6994 & 35.71 & -33.94 &   &    & yes & asterism & {\citet{2001A&A...366..827B}} & {\citet{1971A&A....13..309W}}\\
 & & & & & & &  & {\citet{2000A&A...355..138B}} \\
  & & & & & & &  & {\citet{2002A&A...383..163O}} \\

 & & & & & & &  & {\citet{2007A&A...468..139P}} \\
 & & & & & & &  & {\citet{2018MNRAS.480.5242K}} \\
    \hline    
NGC~7036 & 64.55 & -21.44 & 1000 & 1069 & no & OC & {\citet{2001A&A...366..827B}} & {\citet{1973rncn.book.....S}}\\
 & & & & & & & {\citet{2019A&A...624A...8A}} &  {\citet{2002A&A...385..471C}}   \\
    \hline    
NGC~7055 & 97.45 & 5.62 & 1275 &    & no & OC & {\citet{2012A&A...542A..68P}} & {\citet{1973rncn.book.....S}}\\
    \hline    
NGC~7084 & 69.96 & -24.30 & 765 & 1259 & no & OC & {\citet{2011JKAS...44....1T}} &  {\citet{1973rncn.book.....S}}   \\
 & & & & & & &  & {\citet{2010A&A...516A...3K}} \\
    \hline    
NGC~7127 & 97.90 & 1.14 & 1445 &    & no & OC & {\citet{2011JKAS...44....1T}} & \\
 & & & & & & & {\citet{2012A&A...542A..68P}} &     \\
    \hline    
NGC~7193 & 70.09 & -34.28 & 1080 &    & no & OC & {\citet{2011JKAS...44....1T}} &  {\citet{1973rncn.book.....S}}  \\
 & & & & & & & {\citet{2017RAA....17....4D}} &     \\
 & & & & & & & {\citet{2019A&A...624A...8A}} &     \\
    \hline    
NGC~7772 & 102.74 & -44.27 & 1500 & 1250 & yes & asterism & {\citet{2001A&A...366..827B}} & {\citet{1971A&A....13..309W}}\\
 & & & & & & &   {\citet{2002A&A...385..471C}}  & {\citet{2018MNRAS.480.5242K}} \\
 & & & & & & &   {\citet{2010A&A...516A...3K}}  & {\citet{2019A&A...624A...8A}} \\
    \hline    
NGC~7801 & 114.73 & -11.36 & 1275 & 1953 & no & OC & {\citet{2011JKAS...44....1T}} & {\citet{1973rncn.book.....S}}\\
 & & & & & & & {\citet{2018MNRAS.473..849D}} &     \\
    \hline    
NGC~7826 & 61.87 & -77.65 & 620 &    & no & asterism & {\citet{2011JKAS...44....1T}} & {\citet{1973rncn.book.....S}}\\
 & & & & & & &   &  {\citet{2018MNRAS.480.5242K}}   \\
    \hline   
IC~1023 & 324.95 & 22.71 &    & 1298 & no & OC & {\citet{2001A&A...366..827B}} &    \\
        \hline
Ruprecht~3 & 238.78 & -14.81 & 1100 & 1259 & yes* & asterism & {\citet{2003A&A...399..113P}} & {\citet{2017MNRAS.466..392P}}   \\
 & & & & & & & {\citet{2004A&A...427..485B}} &     \\
 & & & & & & & {\citet{2007A&A...468..139P}} &     \\
 & & & & & & & {\citet{2019A&A...624A...8A}} &     \\
    \hline    
Ruprecht~46 & 238.37 & 5.91 &    & 1467 & yes & OC &     & {\citet{1995MNRAS.276..563C}}\\
    \hline    
Ruprecht~155 & 249.20 & -0.01 & 2311 & 2311 & yes* & OC &     &    \\
    \hline    
Collinder~471 & 110.90 & 13.08 & 2003 & 2210 & yes* & OC &     & {\citet{2018MNRAS.475.4122S}}\\
    \hline    
Basel~5 & 359.77 & -1.87 & 766 & 995 & yes & OC & {\citet{2019MNRAS.488.1635A}} & {\citet{1966ZA.....64...67S}}\\
    \hline    
Loden~1 & 281.02 & -0.17 & 360 & 786 & yes & OC & {\citet{2005A&A...438.1163K}}    & {\citet{2016AJ....152....7H}}\\
    \hline 
    \hline    
        \end{tabular}
\tablefoot{Coordinates from WEBDA when available, else from Simbad. d$_{\mathrm{DAML}}$ and d$_{\mathrm{MWSC}}$: distances listed in \citet{2002A&A...389..871D} and \citet{2013A&A...558A..53K} (respectively). *: WEBDA does not list any parameters other than the sky coordinates.}
\end{center}
\end{scriptsize}
\end{table*}

\section{False positives and confirmation bias} \label{sec:bias}
The fact that we can easily discard as asterisms objects that were, up to now, considered plausible open clusters is largely owed to the spectacular increase in astrometric precision brought by the \textit{Gaia}~DR2 data. With the benefit of hindsight, it is, however, possible to show that some of these objects were always questionable groupings and that the existence of a real cluster was never strongly supported by any data.

In this section we discuss the possible origins of such false positives. Rather than serve merely as a critique of the work published in the literature, these remarks and considerations are aimed at improving the diagnostics and presentation of the results obtained for putative clusters, as there is no doubt the exploitation of the current and upcoming \textit{Gaia} data releases will produce a large number of candidate objects whose nature will not be immediately verifiable.

\subsection{Spatial overdensities}

Human brains have a tendency to seek patterns and are prone to false identifications (e.g. \citealt{rspb.2008.0981}). 
Small-scale areas of relative overdensity always exist in random distributions. 
\citet{2018MNRAS.480.5242K} show that the sparse groupings NGC~1252, NGC~6994, and NGC~7772 (which proved to be asterisms on the basis of proper motions, parallaxes, and radial velocities) only represent a spatial over-density of 1-sigma or less with respect to expected background fluctuations, and that the existence of these clusters was `never very plausible'.

Some studies establish a density profile by binning the data into annuli centred on the apparent location of the density enhancement, and fit a parametrised density model. Fitting a model to binned data is, in fact, not recommended, especially when the underlying data is sparse. \citet{2012arXiv1209.2690T} shows that the result of a fitting procedure to binned data can vary by a surprisingly large amount by simply choosing a different binning\footnote{The same phenomenon at work on a different astronomical problem is mentioned by \citet{2005ApJ...629..873M}, who discuss the bias introduced when determining mass functions from binned data.}. Even the most sincere researcher is likely to choose the arbitrary binning that best illustrates the point they are trying to make and fall victim to confirmation bias.
 Whenever possible (and fitting a density profile is one such case), fitting a model should be done with an unbinned likelihood method. Ideally, the position of the centre itself should be left as a free parameter, as done by \citet{2018MNRAS.477.3600A}. The uncertainties on the best-fit radius and position should also be estimated and reported. Occasionally, some papers contain density profiles that correspond to much less than the 1-sigma uncertainty they display or fit a density profile to stars whose spatial selection was performed manually. In \citet{2007A&A...468..139P}, NGC~1663 stands out from the background by less than one sigma, while NGC~2180 in \citet{2004A&A...427..485B} or NGC~6525 in \citet{2019MNRAS.488.4648P} are indistinguishable from random fluctuations.

\subsection{Signal in photometric space} \label{sec:signal}
The identification of patterns in noisy CMDs can be very subjective. Some groupings had been `confirmed' as clusters based on what the authors interpret as a clear cluster sequence, while other studies estimate that the same CMD contains no visible features. For instance, in two independent investigations of the asterism NGC~6994 (M~73) published almost simultaneously, \citet{2000A&A...355..138B} manually fit a theoretical isochrone to a dispersed distribution of unrelated field stars, while \citet{2000A&A...357..145C} point out the `lack of any feature' in photometric space and concludes that there are `not enough arguments' in favour of the grouping being a physical object.

A procedure that some studies employ in order to extract information from a sparse CMD is to compare it to a nearby offset field \citep[e.g.][]{2007MNRAS.377.1301B,2010MNRAS.407.1875M}. This can be done visually or with an automated de-contamination procedure that removes stars from the investigated CMD based on the photometric structure of the reference field. The interpretation of the results of this procedure is also highly subjective. For instance, some of the offset CMDs shown in \citet{2006JKAS...39..115K} or \citet{2007A&A...468..139P} appear more cluster-like than the central field.

In practice, most studies employ this procedure in situations where the intention is not to clean the cluster CMD, but to `reveal' a cluster sequence that would be invisible otherwise. This approach is undependable for two reasons: i) it does not increase the signal (the cluster sequence) but increases the noise since the Poissonian noise for the background stars in the reference field and cluster field add up and can be quite significant in the low-number count regime; and ii) the result of the subtraction is usually presented as a scatter plot and only shows the areas of the CMD where the cluster field is denser than the reference field. Since the opposite is not shown (areas where the reference field is denser), it is impossible to appreciate the level and structure of the noise, and artefacts can create the illusion of a sequence.

In addition, even in an hypothetical situation where Poissonian noise would be under control, if interstellar extinction is higher around than at the centre (which may be the reason why the asterisms appear as a local enhancement in density in the first place) then subtracting the reddened CMD from the central CMD would create a diagonal artefact that can be mistaken for a cluster sequence (see Fig.~\ref{fig:decontamination}). Although \citet{2007MNRAS.377.1301B} and \citet{2010MNRAS.407.1875M} warn of the limitations of this procedure in case of unknown variable extinction, it is often applied without sufficient justification \citep[e.g. by][]{2018MNRAS.477.3600A,2019MNRAS.488.1635A,2019MNRAS.488.4648P}.

Building a sample of stars selected from their parallax also causes most of the selected stars to align in a sequence in the CMD. Although parallaxes are certainly a valuable piece of information for the selection of cluster members, one should always verify that the cluster is still visible in photometric and proper motion space when the parallax selection is relaxed.

\subsection{Some procedures always return members}

Another situation where confirmation bias plays a significant role is the rejection of outliers. Although discarding data points on the simple basis that they are too discrepant from the bulk of the data (the idea behind the sigma-clipping procedure widely used by astronomers) is a very quick and simple procedure that can produce good results in many situations, it becomes unjusifiable when too many points are removed or when the value expected to be correct is itself poorly defined. 
We refer to \citet{2010arXiv1008.4686H} for a discussion on how to include the modelling of outliers in a fitting procedure without rejecting points a priori. 


Some studies define a membership probability as a distance from an assumed theoretical isochrone \citep[a colour-magnitude filter, e.g.][]{2011JKAS...44....1T}, and proceed to fitting a theoretical isochrone through the non-discarded stars. This approach is seldom justified and it has the untoward effect of allowing one to derive cluster parameters for any initial sample of unrelated stars.

Some membership determination procedures consist of identifying the region of most peaked density in a proper motion diagram. Perhaps the oldest example of this approach can be found in \citet{1958AJ.....63..387V}, who model the proper-motion distribution of NGC~6633 as a mixture of two normal distributions and consider that the component with the smallest variance corresponds to the cluster stars. More sophisticated and non-parametric methods that can be used to separate cluster stars from field stars have been introduced by, for example, \citet{1990A&A...235...94C} and \citet{2016MNRAS.457.3949S}. The drawback of these methods is that they will often converge to a solution and return a sample of `cluster' stars even when the field contains no cluster.

\section{Discussion} \label{sec:discussion}

The aim of this study is not to argue that the only real clusters listed in the catalogues are those that have been identified in the \textit{Gaia}~DR2 data. 
For instance, a notable object absent from our membership list is Saurer~1, one of the most distant clusters currently known, at a distance of $\sim$13\,kpc in the direction of the Galactic anticentre \citep[][]{2004AJ....128.1676C,2006AJ....131..922F,2016A&A...588A.120C}. Although several of its stars are present in the \textit{Gaia}~DR2 catalogue, they are almost all fainter than $G$$\sim$18 and cannot currently be distinguished from the field stars based on astrometric data alone. This object is not, however, controversial\ as multiple authors have obtained comparable results and deep photometry reveals an unequivocal and populated cluster-like sequence. The distant objects identified in the Galactic halo by \citet{2018arXiv181105991P} and \citet{2019MNRAS.484.2181T} also required a combination of \textit{Gaia} data with external photometric catalogues in order to be properly characterised.

On the other hand, the putative object NGC~2234 (not investigated in detail in this study but one that is likely to be an asterism as well) is listed at 6781\,pc in MWSC, but at 4800\,pc in DAML, and as close as 1616\,pc by \citet{2011JKAS...44....1T}, while remaining absent from WEBDA and, additionally, marked non-existent by \citet{1973rngc.book.....S}. For such objects, their alleged large distance is not the most likely explanation for the lack of detection. The large majority of the clusters that have eluded detection in the \textit{Gaia}~DR2 data for no obvious reason have listed distances between 500 and 2500\,pc and they should be considered dubious until they are proven to exist.

A large number of known clusters or candidates listed in the literature were detected by means of infrared photometry and are too obscured to be observed by \textit{Gaia}; for instance, that describes most of the FSR clusters discovered by \citet{2007MNRAS.374..399F} in 2MASS photometry \citep[][]{2006AJ....131.1163S}. Many hundreds of candidate clusters have been identified in infrared surveys, often unresolved or only partially resolved; for instance \citet{2012A&A...542A...3S} \citep[using UKIDSS data,][]{2007MNRAS.379.1599L},    \citet{2013A&A...560A..76M} \citep[using Spitzer/GLIMPSE data,][]{2009PASP..121..213C},  \citet{2018ApJ...856..152R} \citep[using WISE data,][]{2010AJ....140.1868W}, or \citet{2015A&A...581A.120B} \citep[using VVV data,][]{2010NewA...15..433M}. We refer to  the comprehensive work of \citet{2019AJ....157...12B} for an exhaustive compilation of embedded and infrared clusters. Most of these objects will remain forever out of the reach of \textit{Gaia}, but may be characterised one day with data collected by a near-infrared space astrometry mission \citep{2016arXiv160907325H,2019arXiv190705191H,2019arXiv190712535H}.

\subsection{Consequences for the cluster census}

\begin{figure*}[ht]
\begin{center} \includegraphics[scale=0.8]{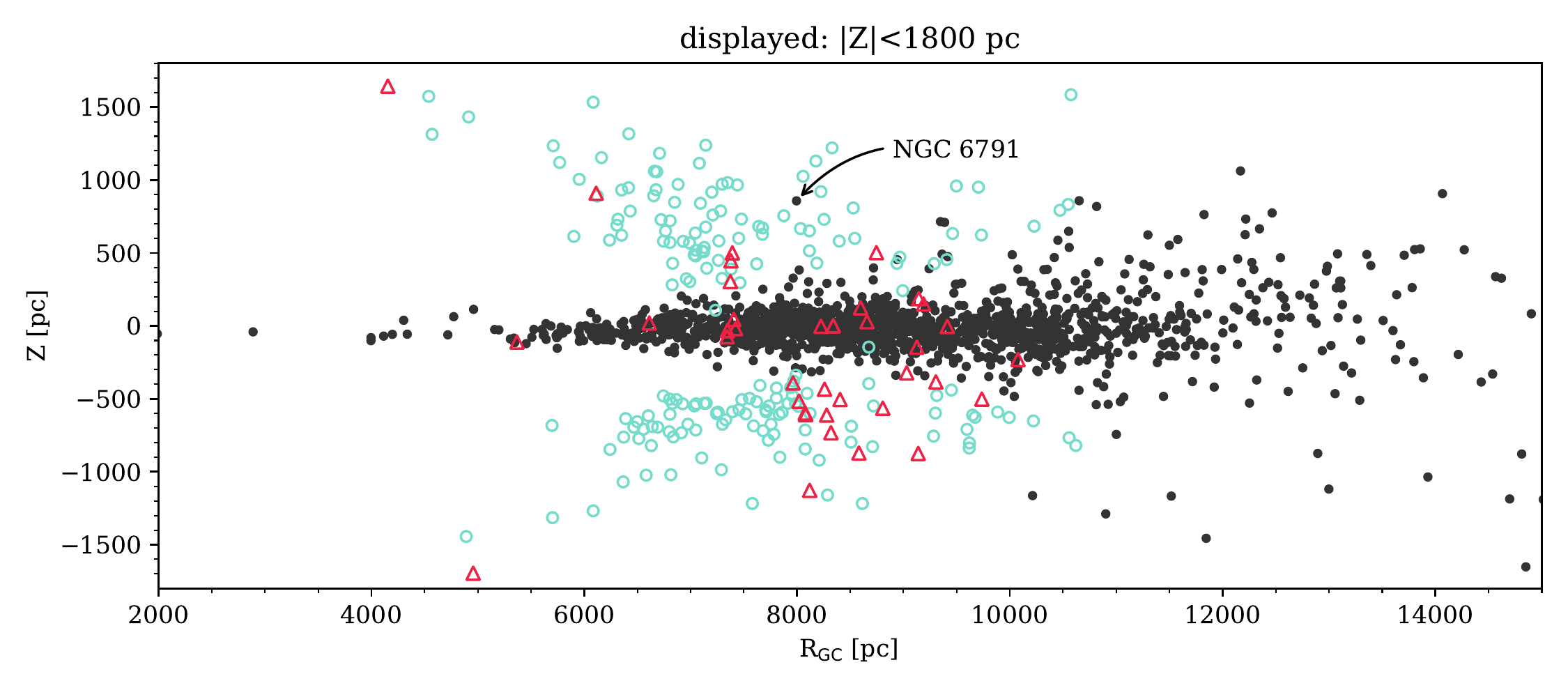} \caption{\label{fig:Rgc_Z_seen_mwsc} Black dots: clusters whose existence has been confirmed with \textit{Gaia}~DR2 data. Open circles: expected location of the candidate clusters of \citet{2014A&A...568A..51S} and \citet{2015A&A...581A..39S}. Open triangles: expected location of the other groupings that this study argues are asterisms. } \end{center}
\end{figure*}

Figure~\ref{fig:Rgc_Z_seen_mwsc} compares the distribution of the detected clusters with those that this study argues are asterisms. Most controversial or erroneous objects mentioned here are alleged old, high-altitude clusters. Their existence would be puzzling and would also have important consequences for our understanding of Galaxy formation and evolution \citep[e.g. the theoretical work of][]{2016ApJ...817L...3M}. Some studies \citep[e.g.][]{2001A&A...366..827B, 2019MNRAS.488.4648P} have proposed that some of these alleged high-altitude objects might belong to the thick disc. 

So far, the only known old high-altitude cluster in the inner disc is NGC~6791 ($z$$\sim$900\,pc). This intriguing old object (7 to 9\,Gyr old, according to \citealt{2005AJ....130..626K,2012A&A...543A.106B}) also features a high metallicity (high-resolution spectroscopy reports {[Fe/H]}=+0.3 to +0.5; e.g.  \citealt{1998ApJ...502L..39P,2006ApJ...643.1151C,2006ApJ...642..462G,2007A&A...473..129C,2009AJ....137.4949B,2012ApJ...756L..40G, 2018AJ....156..142D}), and both its orbital parameters \citep{2012EPJWC..1907005J} and abundance patterns \citep{2014ASPC..482..245C} suggest that it might originate from the Galactic bulge, making it, therefore, non-representative of a hypothetical thick disc cluster population.

The statistical properties of Galactic clusters are often used as probes of the properties of the Galactic disc itself. A number of studies include in their sample the putative clusters of \citet{2014A&A...568A..51S} and \citet{2015A&A...581A..39S}. Along with the 38 asterisms investigated in this study, these false positives collectively amount to 241 non-existing clusters. The studies of \citet{2014MNRAS.444..290B}, \citet{2016A&A...593A.116J}, \citet{2018SSRv..214...74M}, \citet{2018A&A...614A..22P}, \citet{2018sas..conf..216P}, and \citet{2018IAUS..330..227J} also assumed that the cluster census was complete within 1.8\,kpc. Of the 631 confirmed clusters within 1.8\,kpc, 235 were only recently discovered in the \textit{Gaia}~DR2 data. Some of the conclusions of these studies, such as the evolution of the cluster scale height with age, might be quite different in the revised sample. Once the questionable objects are removed, the flaring of the Galactic disc is clearly visible from the distribution of confirmed clusters in Fig.~\ref{fig:Rgc_Z_seen_mwsc} for R$_{\mathrm{GC}}$>10\,kpc. This flaring is visible in Figure~12 of \citet{2018A&A...618A..93C}, who also remark that very few clusters older than $\sim$500\,Myr are known in the inner disc (R$_{\mathrm{GC}}$<7\,kpc). This difference in age distribution between the inner and outer disc seems to indicate that the survival rates of clusters vary significantly with their environment.

 \citet{2018A&A...614A..22P} analysed the age distribution and cluster formation and destruction in the nearby Milky Way disc using the MWSC catalogue. A number of their conclusions are significantly affected by the non-existence of the high-altitude, inner-disc open clusters included in that catalogue. For example, the authors note that the number of evolved clusters had been underestimated in previous results, and that  they  `find an enhanced fraction of older clusters ($t > 1$ Gyr) in the inner disk' but do not observe a `strong variation in the age distribution along [Galactocentric distance]'. These results are clearly the consequences of a contaminated cluster catalogue (see Fig. \ref{fig:Rgc_Z_seen_mwsc}). The derived estimates of the cluster formation and destruction rates, as well as their derived completeness parameters (as well as the cluster age function recently derived from the same data by \citealt{2019ARA&A..57..227K}), would change significantly with our revised sample. In particular, \citet{2018A&A...618A..93C} show that very few old clusters have been confirmed in the inner disc and the cluster survival rates cannot be assumed to be independent of Galactocentric distance. The scale height of several hundreds of parsecs determined by \citet{2016A&A...593A.116J} for the oldest clusters (see their Fig. 5) is also due to the inclusion of non-physical objects: the sample of nearby clusters they used contains 255 objects older than 1\,Gyr, of which only 38 have been recovered with \textit{Gaia}~DR2.

Recent findings have shown that sparse groups of coeval and co-moving stars are not necessarily the remnants of dissolved clusters but may have been sparse since their formation \citep[e.g.][]{2018MNRAS.475.5659W,2019arXiv191006974W}. \citet{2019AJ....158..122K} have identified large-scale co-moving structures that can span over 200\,pc and are not centrally concentrated, but are kinematically cold (with tangential velocity dispersions smaller than a few km\,s$^{-1}$). Some of the groups identified as compact in proper motion space by \citet{2019JKAS...52..145S} are spatially very sparse too (some are in fact so weakly defined spatially that the present study was no able to determine their membership) and so are several of the COIN clusters discovered by \citet{2019A&A...624A.126C}. Studies of the Scorpius-Centauraus \citep{2018MNRAS.476..381W} and Vela-Puppis \citep{2019A&A...626A..17C} stellar complexes have revealed that even very young stellar populations can exhibit sub-structured and non-centrally concentrated spatial distributions spanning hundreds of parsecs and that their overall distribution can reflect the primordial gas distribution rather than the disruption of an initially compact cluster. In this regard, the distinction between clusters and the sparser aggregates traditionally referred to as associations might be arbitrary, with a continuous distribution of possible densities, rather than an objective distinction corresponding to fundamentally different formation mechanisms \citep[e.g.][]{2019arXiv190709712P}. Therefore, we argue that classifying an object as a remnant should not be done on the basis of morphological properties but should  be based on further physical arguments, such as an evident deficit of low-mass stars, as in e.g. NGC~7762 \citep{1995A&AS..114..281P} or Ruprecht~147 \citep{2019AJ....157..115Y}. The clusters NGC~1901, NGC~3680 (discussed in Sect.~\ref{sec:intrinsic}), along with NGC~7762 and Ruprecht~147, can be considered good examples of the late dynamical stage of a stellar cluster.

\subsection{Good practice}
Some Galactic clusters have been the subject of a large number of studies, while others are hardly ever mentioned in the literature. Investigating the properties of the neglected objects is a laudable, useful, and fulfilling endeavour. In some cases, the `poorly studied' or `hitherto unstudied' cluster does not have its parameters given in the literature but may be mentioned by authors who reportedly failed to identify it or explicitly propose that the cluster does not exist. Such comments are, however, not always available as researchers are more likely to report on their successes than their failures. Since science is a process fueled by unsuccessful attempts and failed experiments, it might be a good habit for papers presenting cluster searches to name the objects for which the search was unsuccessful \citep[as done by e.g.][]{1971A&AS....4..241B,2019MNRAS.487.2385M}. 



Studies based on \textit{Gaia} data allow us to verify that the proper-motion and parallax dispersion of a group of stars is indeed compatible with them forming a cluster. Parallaxes also make it easy to verify if the distance modulus estimated from photometry agrees with the \textit{Gaia} measurements. For several objects mentioned in this study \citep[and in][]{2018MNRAS.480.5242K}, the cluster members proposed by various authors are entirely different groups of stars. It is therefore important for the reproducibility of the results that membership lists are published (in electronic form) along with the papers. This also makes it easier to verify the properties of a group of stars when new data is available.

The quality of \textit{Gaia} being superior to that of early 19th century instruments, it sounds unlikely that objects discovered by visual observers might be difficult to find for modern astronomers. In the collective endeavour of charting the Milky Way, we should therefore trust the current data rather than the old catalogues. In the words of \citet{2018MNRAS.480.5242K}, `the existence of sparse clusters should be double-checked, regardless of how reputable the respective cluster catalogues are'. In this regard, \citet{2010A&A...516A...3K} mention that `to avoid any prejudice', they do not display the cluster names on the figures when inspecting proper motion diagrams and DSS images.


\begin{figure*}[ht!]
\begin{center}
\includegraphics[width=0.9\textwidth]{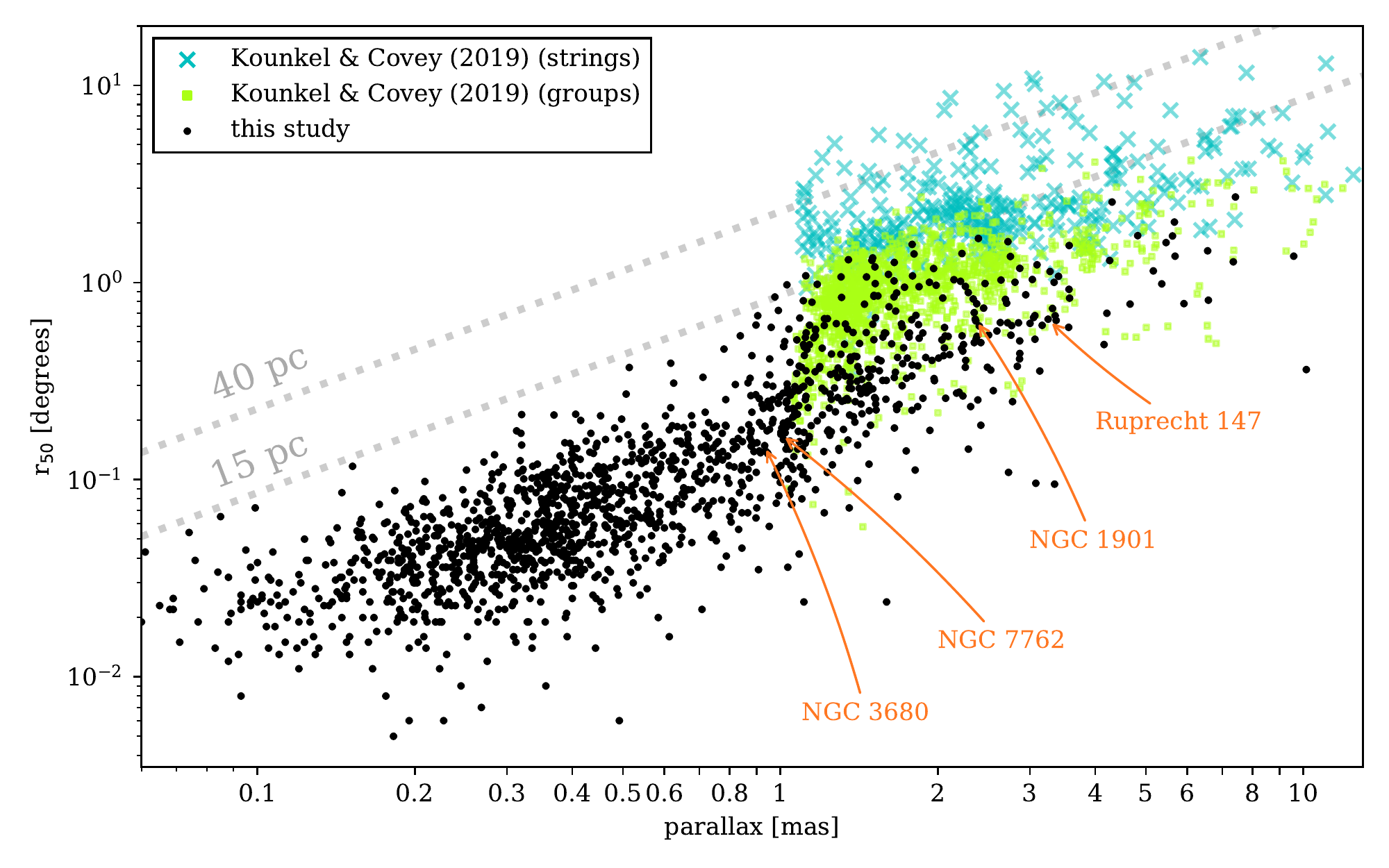} 
\caption{Apparent radius against mean parallax for the clusters confirmed by {\it Gaia}~DR2 as of this paper (black dots) as well as the isolated groups (green) and elongated structures belonging to "strings" (cyan crosses) reported by \citet{2019AJ....158..122K} in the \textit{Gaia}~DR2 data. For the latter data, a proxy for $r_{50}$ was estimated as $0.5 \times \sqrt{{\rm width}^2+{\rm height}^2}$. Four dynamically evolved clusters are labelled. The dotted lines indicate the angular size corresponding to 15 and 40\,pc.} \label{fig:r50_vs_par}
\end{center}
\end{figure*}

\subsection{Empirical criteria for bona fide clusters with {\it Gaia}} \label{sec:minimal}

We propose a set of simple, observationally-motivated criteria that can be applied to assess the reality of dubious objects with {\it Gaia} data. This empirical set of conditions is not a rigorous physical modelling \citep[where considerations on stellar dynamics would require well-resolved kinematics or knowledge of the total mass as in e.g.][]{2011MNRAS.410L...6G} but meant as a guideline for discarding implausible objects.

{\it Total proper-motion dispersion:} The conservative velocity dispersion upper limit of 5~km\,s$^{-1}$ presented in Sect.~\ref{sec:intrinsic} and illustrated in Fig.~\ref{fig:totPMdisp_vs_par} translates into a total proper-motion\footnote{We recall: proper motion $\mu \simeq v \varpi / 4.74$, expressed in mas\,yr$^{-1}$   if velocity $v$ in km\,s$^{-1}$ and parallax $\varpi$ in mas.} dispersion as: 
    
    \begin{equation*}
    \sqrt{\sigma_{\mu*_{\rm \alpha}}^2+\sigma_{\mu_{\rm \delta}}^2} \lesssim  
  \begin{cases}
    1~ {\rm mas\,yr^{-1}} & \text{if } \varpi \leq 1~ {\rm mas} \\
    5 \sqrt{2}  \frac{\varpi}{4.74} ~ {\rm mas\,yr^{-1}} & \text{if } \varpi > 1~ {\rm mas} \\
  \end{cases} 
  \end{equation*}
  
  This condition would, in fact, discard the most massive globular clusters if they were located closer than 1\,kpc from us. Such objects would, however, be extremely obvious in the night sky, and proving their existence would not require {\it Gaia} astrometric data.

{\it Sky concentration:} Known clusters span a wide range of masses and physical sizes. 
    In Fig.~\ref{fig:r50_vs_par} we show the sky concentration $r_{50}$ in degrees (defined as the radius in which half of the identified members are located) as a function of mean cluster parallax, for clusters confirmed with {\it Gaia}~DR2. Very few of them exhibit angular sizes corresponding to physical dimensions beyond 15\,pc. The study of \citet{2019AJ....158..122K} has identified sparse and elongated structures that can have characteristic sizes of several tens of parsecs. Most of them are groups of young stars tracing the original gas distribution in their parent molecular clouds, and are not necessarily gravitationally-bound. Despite being old and dynamically-evolved, the four clusters labelled in Fig.~\ref{fig:r50_vs_par} are not physically very extended. Therefore, the unusually large spatial extension of some putative clusters cannot simply be explained by them being cluster remnants.

{\it Parallax distribution:} The intrinsic parallax dispersion of cluster members must correspond to a physically plausible depth. For distant clusters, the parallax distribution of members is dominated by errors, and the individual parallaxes must be compatible with being drawn from the same true parallax. One possible way to estimate the intrinsic parallax dispersion for a set of sources is to perform a maximum likelihood estimation that assumes a normal distribution and takes into account individual parallax uncertainties. 
    
    For the proposed members of the asterism NGC~1663 (Sect.~\ref{sec:NGC_1663}), we recover an intrinsic parallax dispersion of 0.15$^{+0.09}_{-0.05}$\,mas and a mean of 0.36$\pm$0.10\,mas, which corresponds to an unphysical physical dispersion of several kiloparsecs along the line of sight.

{\it Colour-magnitude diagram:} The colour-magnitude diagram of any physical open cluster should follow an empirical isochrone, convolved with typical measurement errors, and possibly blurred by interstellar extinction. This requirement cannot easily be transformed into a mathematical criterion as it requires visual inspection by experts or well-trained machine-learning algorithms.

{\it Minimum number of stars:} A commonly used minimum is ten stars \citep[e.g.][]{2018A&A...618A..59C}, or slightly less \citep{2019JKAS...52..145S}. The number of identified cluster members may depend not only on the cluster itself, but also on its distance, age, velocity relative to the field stars, and density of the background stellar distribution. Cluster candidates with a dozen or fewer proposed members should be considered dubious unless they can be shown to clearly pass all of the above conditions.

\section{Summary and conclusion} \label{sec:conclusion}

In this study, we derive lists of cluster members for objects that were not included in \citet{2018A&A...618A..93C} either because they were not identified or because they had not yet been discovered. We bring the total number of clusters with available membership from \textit{Gaia}~DR2 to 1481. We also investigate 38 objects whose trace is not visible in the \textit{Gaia}~DR2 astrometry and we argue that they are not real clusters. Many of them have been flagged as asterisms or non-existent in one or in multiple catalogues but are still included in recent studies (in particular those that were believed to be old, high-altitude, inner-disc remnants of open clusters). 

Since its release, the \textit{Gaia}~DR2 data has shown that about a third of the proposed open clusters listed in the catalogues within 2\,kpc are not true clusters. A roughly similar number of new clusters have been discovered. The Milky Way disc traced by the objects that we do detect in \textit{Gaia}~DR2 data shows a clear lack of both old and high-altitude clusters in the inner regions. Although the census might still be affected by observational biases (because detecting objects against the crowded background of the inner Milky Way might be more difficult), this distribution strongly supports the idea that the time scale for destruction is faster in the inner disc, and clusters do not have time to migrate to high altitudes before being destroyed.

The tale of the sparse NGC clusters, originally reported by visual observers in the 18th and 19th century, and whose names and coordinates were carefully passed on from paper to electronic catalogues without any tangible proof of their existence, bears resemblance to phantom islands, misreported lands that were copied down by cartographers (sometimes for centuries) until enough evidence was collected to disprove their existence. This study is far from having investigated all the known clusters that have not yet been identified in \textit{Gaia}~DR2 data and the Galactic cluster catalogues likely need further cleaning. In particular, objects that have not been detected with {\it Gaia}~DR2 but cannot currently be proven to be asterisms should be re-investigated with the upcoming \textit{Gaia} data releases.

With near-infrared (NIR) space astrometry coming within reach \citep{2012ASPC..458..417G, 2019BAAS...51c.118M, 2019arXiv190705191H}, a NIR version of the {\it Gaia} satellite may soon become viable. A {\it GaiaNIR}-like mission \citep{2019arXiv190712535H} will be transformative for studies of infrared clusters, in a similar way as {\it Gaia} is currently transforming our optical view of the cluster population of the Milky Way. The principles of good practice for an astrometric census of star clusters discussed in this paper will then be applicable to infrared data.


\section*{Acknowledgements}
The authors thank C. Jordi for her feedback on the manuscript, and the anonymous referee for a thorough report that greatly helped to improve the quality of this paper.

This work has made use of data from the European Space Agency (ESA) mission \textit{Gaia} (www.cosmos.esa.int/gaia), processed by the \textit{Gaia} Data Processing and Analysis Consortium (DPAC, www.cosmos.esa.int/web/gaia/dpac/consortium). Funding for the DPAC has been provided by national institutions, in particular the institutions participating in the \textit{Gaia} Multilateral Agreement. 
This work was supported by the MINECO (Spanish Ministry of Economy) through grant ESP2016-80079-C2-1-R and RTI2018-095076-B-C21 (MINECO/FEDER, UE), and MDM-2014-0369 of ICCUB (Unidad de Excelencia 'María de Maeztu'). TCG acknowledges support from Juan de la Cierva - Formaci\'on 2015 grant, MINECO (FEDER/UE). FA is grateful for funding from the European Union's Horizon 2020 research and innovation programme under the Marie Sk\l{}odowska-Curie grant agreement No. 800502.

The preparation of this work has made extensive use of Topcat \citep{Taylor05}, and of NASA's Astrophysics Data System Bibliographic Services, as well as the open-source Python packages \texttt{Astropy} \citep{Astropy13}, \texttt{NumPy} \citep{VanDerWalt11}, \texttt{scikit-learn} \citep{scikit-learn}, and \texttt{APLpy}.
The figures in this paper were produced with Matplotlib \citep{Hunter07}. This research has made use of Aladin sky atlas developed at CDS, Strasbourg Observatory, France \citep{Bonnarel00aladin,Boch14aladin}. 

\bibliographystyle{aa} 
\linespread{1.5}                
\bibliography{clusters_biblio}

\appendix

\section{Notes on individual asterisms}\label{sec:indiv_asterisms}
\subsection{NGC~1252} \label{indiv:start}

\citet{1973rncn.book.....S} marked this object as unverified in the Revised NGC. \citet{1983PASP...95..474B} mention that it is `difficult to decide whether the cluster is real' but also list 14 possible members. \citet{1984PASP...96...70E} considered the existence of this object unlikely and \citet{1998A&A...340..402B} made use of Hipparcos data to conclude that NGC~1252 does not exist. The asterism is listed as a potential remnant in \citet{2001A&A...366..827B} and investigated in detail by \citet{2001A&A...374..554P}, who estimate a distance of $\sim$640\,pc and an age of 3\,Gyr on the basis of a very sparse and noisy CMD. The object is included in the DAML and MWSC catalogues (with quoted distances of 790 and 944\,pc, respectively) and is also present in the study of \citet{2007A&A...468..139P}.

\citet{2013MNRAS.434..194D} collected additional data and reported that most stars in the investigated region are `chemically, kinematically, and spatially unrelated to each other' but still argue that a handful of faint stars might be co-moving, making this `enigmatic object' the first old, high-altitude ($b=-50.8^{\circ}$), nearby open cluster. \citet{2013A&A...558A..53K} also provide the result of an isochrone fitting procedure for NGC~1252, indicating an old age of $\log t$=9.5

\citet{2018A&A...618A..93C} were unable to identify any such co-moving group in the \textit{Gaia}~DR2 catalogue, and \citet{2018MNRAS.480.5242K} also failed to identify one despite complementing the \textit{Gaia}~DR2 data with new high-resolution spectroscopy. They also point out that not a single pair of stars in the entire member lists of \citet{1983PASP...95..474B}, \citet{2001A&A...374..554P}, \citet{2013MNRAS.434..194D}, and \citet{2013A&A...558A..53K} have matching astrometric parameters. NGC~1252 is listed as an asterism in \citet{2019AJ....157...12B} and \citet{2019A&A...624A...8A}.

\subsection{NGC~1520} 
This object is not listed in the WEBDA database and is flagged as non-existent in the revised NGC catalogue \citep{1973rncn.book.....S}. It is included in the DAML catalogue with a distance of 775\,pc, and in the MWSC catalogue at 1023\,pc, with ages of $\log t$=9.3 and 9.43 (respectively).
\citet{2019AJ....157...12B} list it as a cluster remnant.

\subsection{NGC~1557} 
This object is not included in WEBDA and is classified as non-existent by \citet{1973rncn.book.....S}, but is listed as a cluster remnant by \citet{2001A&A...366..827B} and \citet{2019AJ....157...12B}. 
\citet{2011JKAS...44....1T} estimate an age $\log t$=9.48 and a distance of 1055\,pc (these numbers are quoted in the DAML catalogue), while MWSC lists an age of 9.5 and a distance of 1820\,pc.

\subsection{NGC~1641} 

This compact and irregular grouping located at $(\ell,b)$=(277.20$^{\circ}$,-38.32$^{\circ}$) was considered by \citet{1963IrAJ....6...74S} \citep[and later by][]{1973rncn.book.....S} to be related to in the Large Magellanic Cloud \citep[although][remark its irregular shape]{1963IrAJ....6...74S}. It is considered a remnant of a Milky Way open cluster by \citet{2001A&A...366..827B}. It was studied by \citet{2006JKAS...39..115K}, who determined a distance of 1.2\,kpc and an age of 1.6\,Gyr by manually fitting theoretical isochrones to a very sparse and noisy CMD. 
\citet{2019AJ....157...12B} also flag NGC~1641 as an open cluster remnant. \citet{2002A&A...389..871D} and \citet{2013A&A...558A..53K} both quote a distance of 985\,pc and an age of $\log t$=9.52.

\subsection{NGC~1663} \label{sec:NGC_1663}
This grouping, described as `not rich' in J.~Herschel's original notes, was studied by \citet{2003A&A...407..527B}, who concluded that `it is hard to decide upon the real nature of this cluster'. The DAML catalogue reports a distance of 700\,pc and MWSC reports a distance of 1490\,pc (they report values of $\log t$ 9.3 and 9.4, respectively). This asterism is considered a remnant by \citet{2007A&A...468..139P} and  \citet{2019AJ....157...12B}.

\begin{figure}[ht]
\begin{center} \resizebox{0.49\textwidth}{!}{\includegraphics[scale=0.6]{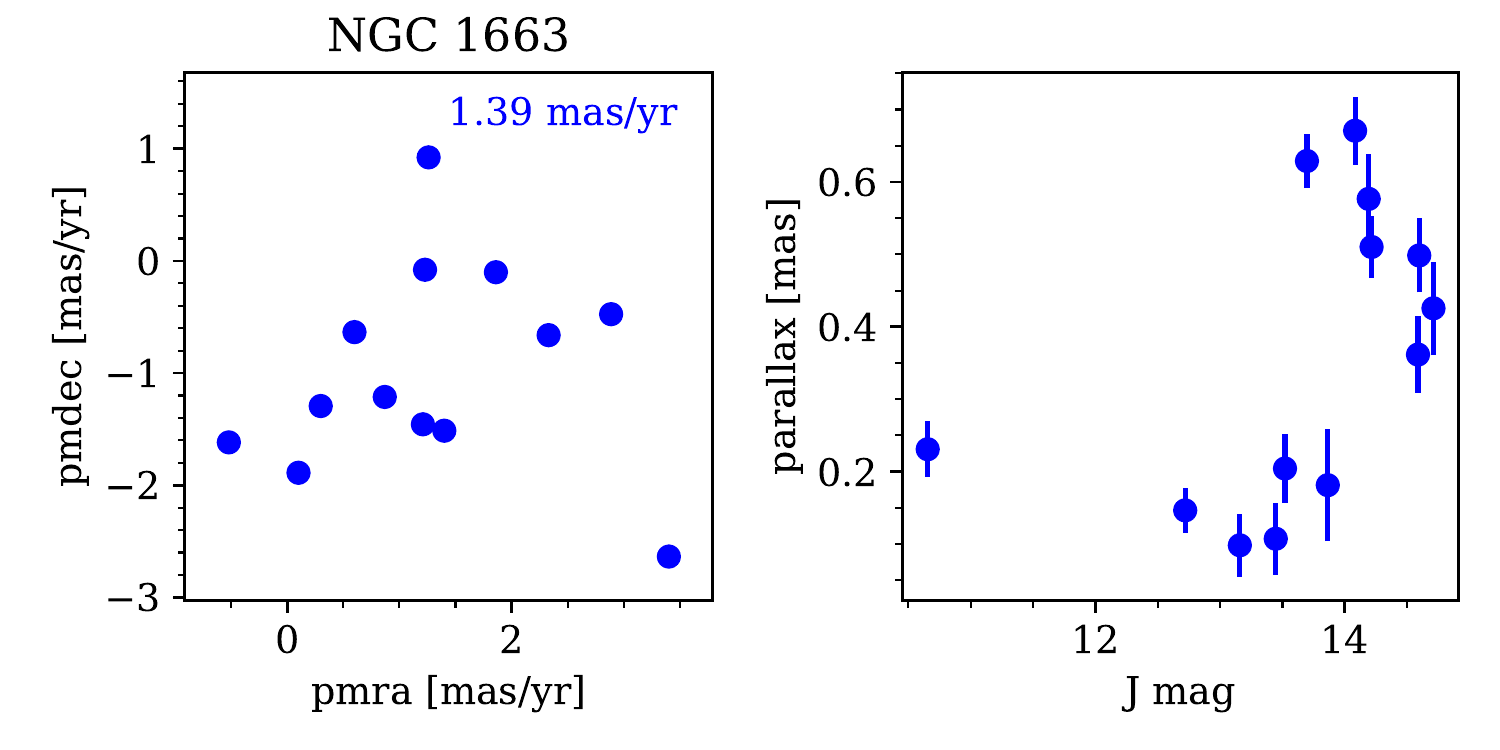}} \caption{\label{fig:angelo19members_NGC_1663} NGC~1663. Left: \textit{Gaia}~DR2 proper motions for the members proposed by \citet{2019A&A...624A...8A}. The error bars are smaller than the markers. The total proper-motion dispersion is indicated. Right: parallax vs 2MASS $J$ mag for the same stars. } \end{center}
\end{figure}

This object was also studied by \citet{2019A&A...624A...8A}, who propose 13 possible members. Figure~\ref{fig:angelo19members_NGC_1663} shows that these 13 stars do not form a coherent group in either \textit{Gaia}~DR2 proper motion or parallax space. Assuming that all 13 stars were drawn from a single normal distribution, a maximum likelihood estimation yields a mean of 0.36$\pm$0.10\,mas and an intrinsic dispersion of 0.15$^{+0.09}_{-0.05}$\,mas (corresponding to a distance dispersion of 2 to 3\,kpc), which shows that the parallax scatter cannot be explained by astrometric errors.

\subsection{NGC~1746} 
This group of stars was entered in the NGC catalogue as a sparse distribution overlapping with the more compact NGC~1750 and NGC~1758. Later references sometimes considered all three to be one single object, catalogued as NGC~1746 \citep[e.g.][]{1937AnHar.105..403C,2013A&A...558A..53K}. All three objects are listed as open clusters in {\citet{1973rncn.book.....S}}.

 {\citet{1992BaltA...1..125S}}, {\citet{1998A&A...333..471G}},  {\citet{1998A&AS..131...89T}}, and {\citet{2010PASP..122.1008L}} have reportedly identified NGC~1750 and NGC~1758 as two distinct groups, but do not find the trace of a third object that could be identified as NGC~1746. \citet{2018A&A...618A..93C} reached the same conclusion with \textit{Gaia}~DR2 data. NGC~1746 is however still listed as an open cluster in \citet{2019AJ....157...12B}.

\subsection{NGC~1963}
This object is not listed in WEBDA and does not seem to have ever been the subject of a dedicated study before \citet{2001A&A...366..827B}. \citet{1973rncn.book.....S} mark it as non-existent.
It is not listed in DAML, but is in MWSC with a distance of 1700\,pc and $\log t$=8.125. \citet{2011JKAS...44....1T} mention that they were unable to find the trace of this cluster in positional or photometric space.

\subsection{NGC~2132}
This object is not listed in WEBDA and does not seem to have ever been the subject of a dedicated study. \citet{1973rncn.book.....S} classify it as non-existent.
The DAML catalogue reports a distance of 974\,pc, and MWSC a distance of 1000\,pc (with $\log t$ of 9.22 and 9.045, respectively).
This asterism is flagged as a remnant in \citet{2019AJ....157...12B}.

\subsection{NGC~2180}

This object was first reported by W. Herschel. \citet{1973rncn.book.....S} flag it as non-existent in the Revised NGC. 
The catalogues of DAML and MWSC quote discrepant distances of 910 and 1882\,pc.

\begin{figure}[ht]
\begin{center} \resizebox{0.49\textwidth}{!}{\includegraphics[scale=0.6]{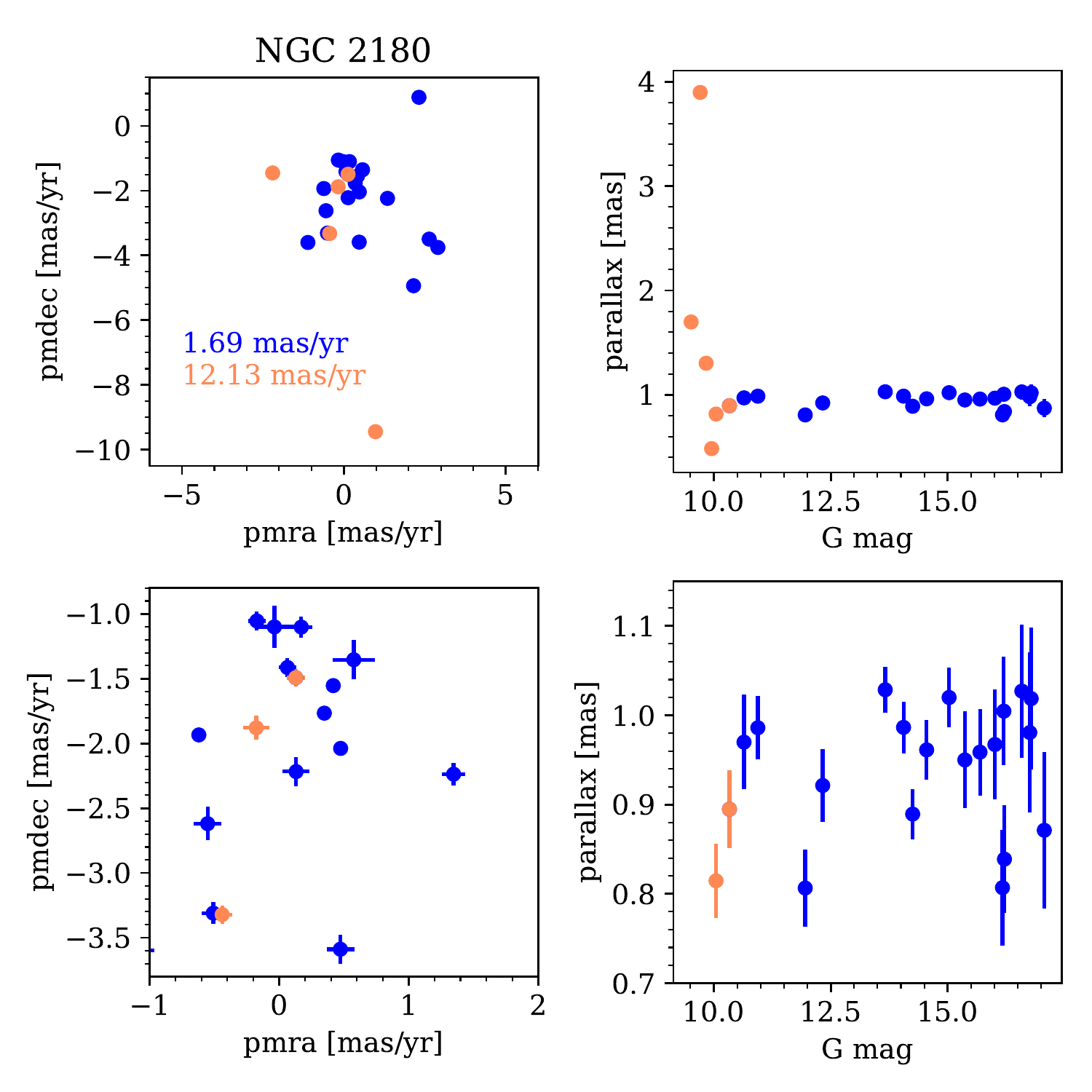}} \caption{\label{fig:angelo19members_NGC_2180} NGC~2180. Top left: \textit{Gaia}~DR2 proper motions for the members proposed by \citet{2019A&A...624A...8A} (blue) and five of the six members proposed by \citet{2004A&A...427..485B} (orange). The sixth member is outside the range of the plot. The error bars are smaller than the markers. The total proper-motion dispersion is indicated for both samples. Top right: parallax vs $G$ mag for the same stars. Bottom row: same as top row, with a different scale.} \end{center}
\end{figure}

The object was investigated by \citet{2004A&A...427..485B}, who estimate a distance of $\sim$900\,pc and an age of 710\,Myr from a putative cluster sequence in a CMD, and propose that the object is a dissolving open cluster. They identify six potential red clump stars. Figure~\ref{fig:angelo19members_NGC_2180} shows that the \textit{Gaia}~DR2 proper motion and parallax of these stars are inconsistent with them forming a cluster.

\citet{2011MNRAS.412.1611P} used NGC~2180 as a reference object, to establish the physical reality of other candidate groupings. \citet{2008A&A...477..165P} also provide a mass estimate for this object.

\citet{2019A&A...624A...8A} report 20 members, with only one in common with \citet{2004A&A...427..485B}. The proper-motion distribution of these stars does not form a coherent cluster either (Fig.~\ref{fig:angelo19members_NGC_2180}).

\subsection{NGC~2220} 
This grouping was originally described as `poor, very coarsely scattered' by J. Herschel. The Revised NGC of \citet{1973rncn.book.....S} flagged it as non-existent and the object is not present in the WEBDA database, but it is included in the DAML catalogue (1170\,pc, $\log t$=9.48) and MWSC (1393\,pc, $\log t$=9.68). This asterism is flagged as a remnant in \citet{2019AJ....157...12B}.

\subsection{NGC~2348} 
Originally reported as a `coarse loose cluster' by J.~Herschel, this entry was flagged as unverified in the Revised NGC of \citet{1973rncn.book.....S} and is not present in the WEBDA database. The catalogues of DAML and MWSC quote distances of 1070 and 1076\,pc, and $\log t$=9.26 and 9.475, respectively. This object is listed as a cluster remnant in \citet{2001A&A...366..827B} and \citet{2019AJ....157...12B}.

\subsection{NGC~2394} 

Reported by J.~Herschel as `very coarsely scattered' and `not rich', this entry was marked as non-existent in the Revised NGC of \citet{1973rncn.book.....S}. \citet{2006JKAS...39..115K} estimates a distance of 660\,pc and an age of 1.1\,Gyr, while DAML quote 940\,pc and $\log t$=8.95. This object is not present in the MWSC catalogue and is flagged as an asterism in \citet{2019AJ....157...12B}.

\subsection{NGC~3231}
This entry was flagged as non-existent in the Revised NGC catalogue \citep{1973rncn.book.....S} and is not included in the WEBDA database nor in the MWSC catalogue but is in the DAML catalogue who quotes the distance of 715\,pc and the age of 7\,Gyr estimated by \citet{2011JKAS...44....1T}. \citet{2012A&A...542A..68P} report that they could not identify any kinematic overdensity corresponding to this object in the PPMXL data and suggest it could be classified as a remnant.

\begin{figure}[ht]
\begin{center} \resizebox{0.49\textwidth}{!}{\includegraphics[scale=0.6]{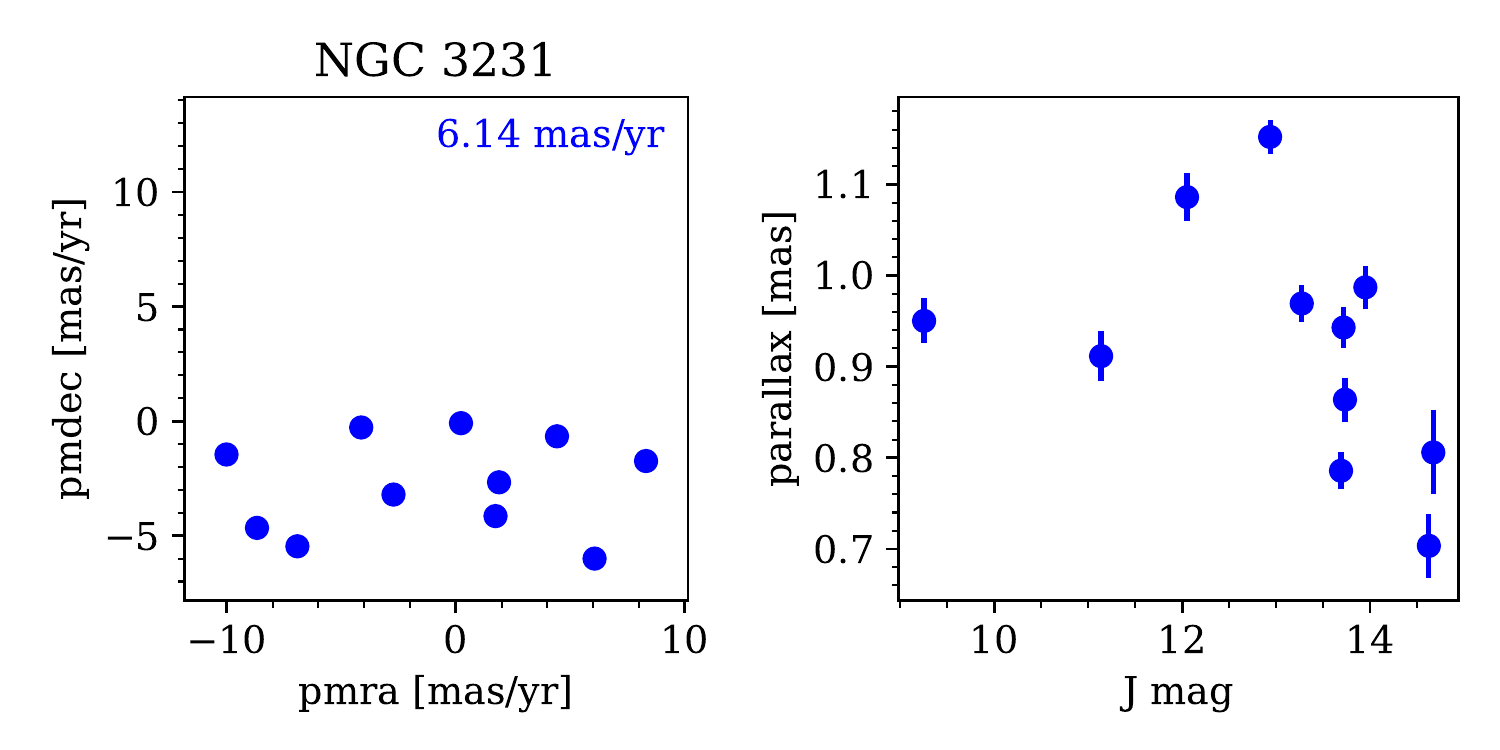}} \caption{\label{fig:angelo19members_NGC_3231} NGC~3231. Left: \textit{Gaia}~DR2 proper motions for the members proposed by \citet{2019A&A...624A...8A}. The error bars are smaller than the markers. The total proper-motion dispersion is indicated. Right: parallax vs 2MASS $J$ mag for the same stars. } \end{center}
\end{figure}

This object was also studied by \citet{2019A&A...624A...8A}, who propose 11 possible members. Figure~\ref{fig:angelo19members_NGC_3231} shows that these stars do not form a coherent group in either proper motion or parallax space.

\subsection{NGC~4230}
This object is included in the WEBDA database but with no listed parameters. The DAML quotes a distance of 1445\,pc and $\log t$=9.23 \citep[taken from][]{2011JKAS...44....1T}, while MWSC quotes 2630\,pc and $\log t$=8.9. \citet{2019MNRAS.488.4648P} report a distance modulus corresponding to 3470\,pc and a spatial density profile compatible with random fluctuation.

\subsection{NGC~5269}
John Herschel originally reported this grouping as `poor, large, loose, irregular'. The entry was marked non-existent by \citet{1973rncn.book.....S}. The object is not present in WEBDA and \citet{2011JKAS...44....1T} reported that they were unable to identify the trace of a cluster. NGC~5269 is however in DAML (1410\,pc, $\log t$=8.2) and in MWSC (1634\,pc, $\log t$=8.52).

Recently, \citet{2017MNRAS.466.4960P} identified four stars with matching proper motions and parallaxes in the \textit{Gaia}~DR1 catalogue and proposed a distance of 2\,kpc and an age of $\log t$=8.5.

\subsection{NGC~5998}
This object was marked non-existent by \citet{1973rncn.book.....S} and is not present in the WEBDA database. The catalogues of DAML and MWSC quote very different distances (1170 and 4853\,pc, respectively) and $\log t$ of 9.2 and 9.5. \citet{2011JKAS...44....1T} reports a distance of 981\,pc.

\subsection{NGC~6169}

This object has an entry in WEBDA but no listed parameters. \citet{1973A&AS...10..135M} and \citet{2011JKAS...44....1T} report that they were unable to find a trace of this object, while DAML and MWSC both quote a distance of 1007\,pc and $\log t$=7.5.

\subsection{NGC~6481}
This object was marked non-existent by \citet{1973rncn.book.....S} and is not present in the WEBDA database. The catalogues of DAML and MWSC report distances of 1180 and 1836\,pc, and $\log t$=9.5. \citet{2007A&A...468..139P} and \citet{2019A&A...624A...8A} consider it an open cluster remnant.

\begin{figure}[ht]
\begin{center} \resizebox{0.49\textwidth}{!}{\includegraphics[scale=0.6]{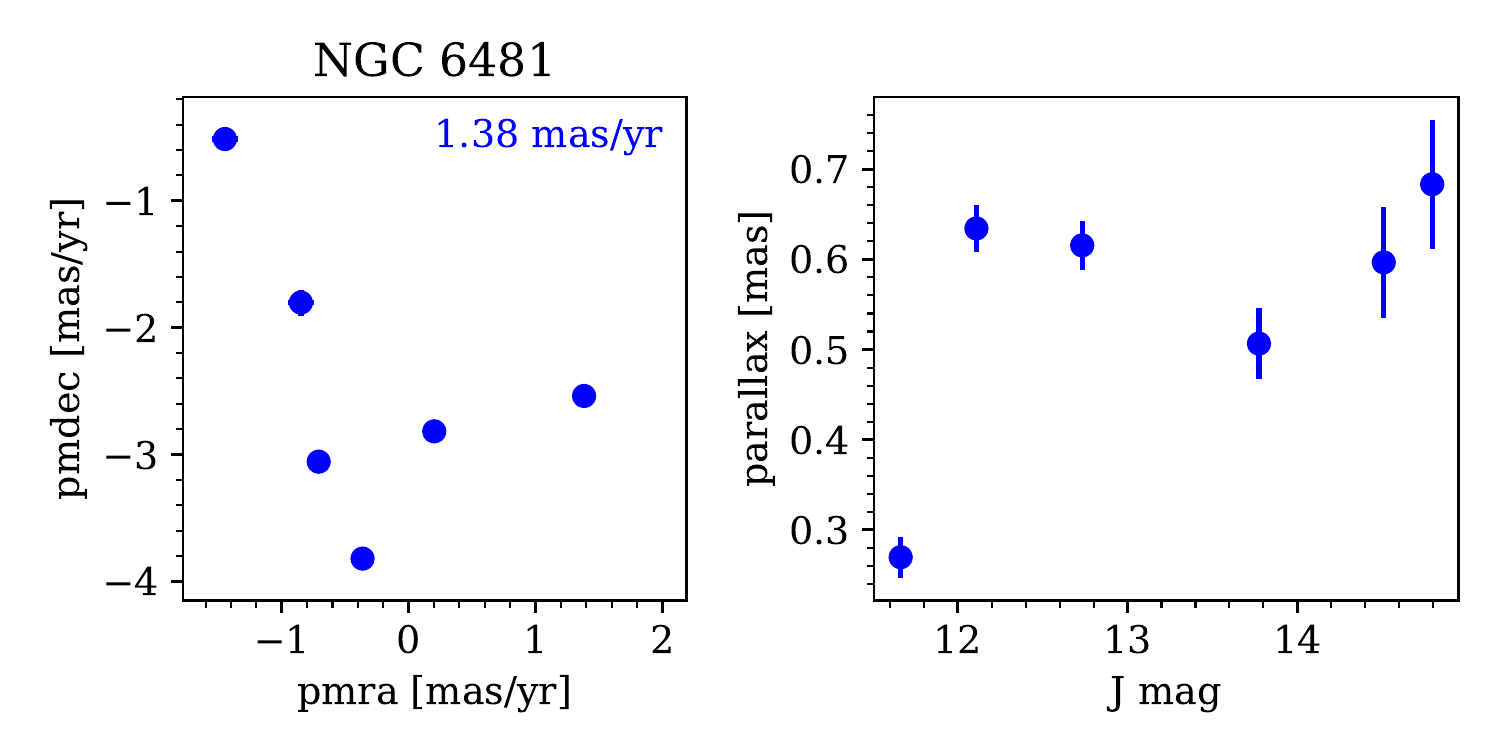}} \caption{\label{fig:angelo19members_NGC_6481} NGC~6481. Left: \textit{Gaia}~DR2 proper motions for the members proposed by \citet{2019A&A...624A...8A}. The error bars are smaller than the markers. The total proper-motion dispersion is indicated. Right: parallax vs 2MASS $J$ mag for the same stars. } \end{center}
\end{figure}

\citet{2019A&A...624A...8A} propose 6 possible members. Figure~\ref{fig:angelo19members_NGC_6481} shows that these stars do not form a coherent group in either proper motion or parallax space.

\subsection{NGC~6525}
This object was marked non-existent by \citet{1973rncn.book.....S} and is not present in the WEBDA database. The catalogue of DAML quote a distance of 1436\,pc and age of 2\,Gyr \citep[after][]{2011JKAS...44....1T}, while MWSC quotes 3221\,pc and $\log t$=9.45. The MWSC catalogue however flags it as not found in DSS images inspection. 

\citet{2010A&A...516A...3K} report that attempting to estimate a collective proper motion for this object led to a poor solution and recall that it is one of the non-existent clusters of \citet{1973rncn.book.....S}. 

\citet{2019MNRAS.488.4648P} consider this object an open cluster remnant at a distance of 3300\,pc, although the constructed density profile shows an overdensity of less than one sigma in significance and a proper-motion dispersion that appears close to $\sim$10\,mas\,yr$^{-1}$.

\subsection{NGC~6554}
This object was marked non-existent by \citet{1973rncn.book.....S} and is not present in the WEBDA database. \citet{2011JKAS...44....1T} report that they looked for this object and were unable to find it. The catalogues of DAML and MWSC both quote a distance of 1775\,pc and $\log t$=8.62.

\subsection{NGC~6573}
This object was marked non-existent by \citet{1973rncn.book.....S} and is not present in the WEBDA database, but is present in DAML (listed at a distance of 460\,pc and age $\log t$=7) and in MWSC (at a distance of $\sim$3000\,pc and age $\log t$=8.8, flagged as an open cluster remnant). It is included in the list of open cluster remnants of \citet{2018MNRAS.477.3600A}.

\subsection{NGC~6588}
This object was marked non-existent by \citet{1973rncn.book.....S} and is not present in the WEBDA database. The distance estimates available in the literature are very discrepant: \citet{2011JKAS...44....1T} estimate a distance of 960\,pc (and an age of 1.6\,Gyr), while DAML quote 2314\,pc ($\log t$=9.65) and MWSC quotes 4757\,pc ($\log t$=9.4). 

\citet{2015NewA...38...31C} and \citet{2017NewA...51...15M} derive distances of 2314\,pc and 2159\,pc (respectively) and an age $\log t$=9.5, from $UBVRI$ photometry.

\subsection{NGC~6994 (Messier~73)}

Originally reported in Messier's catalogue as a group of `three or four small stars' potentially surrounded by a nebula, this asterism was listed in J. Herschel's General Catalogue under entry GC~4617, and flagged as `Cl??; eP; vlC; no neb', standing for: cluster of very doubtful existence; extremely poor; very little concentrated; no nebulosity \citep{1864RSPT..154....1H}. This asterism still made its way into Dreyer's New General Catalogue \citep{1888MmRAS..49....1D}, who copied Herschel's notes without including the question marks. \citet{1908AnHar..60..199B} lists it as a ``coarse cluster''. Due to the small angular size of the asterism on the sky, \citet{1931AnLun...2....1C} estimated a distance of 14270 light years (4275\,pc). The object is listed by \citet{1958csca.book.....A} and included in the catalogue of \citet{1966BAICz..17...33R} as a class `IV 1 p' (sparse and poor) open cluster. \citet{1968ArA.....5....1L} do not list it in their catalogue of cluster ages, and \citet{1971A&A....13..309W} mention it as a doubtful cluster, but the object is still listed by \citet{1995yCat.7092....0L}.

\citet{2000A&A...355..138B} performed the first photometric study of NGC~6994, identified 24 members, and fit an isochrone to a $BV$ CMD to find a distance of 620\,pc \citep[at odds with the 4275\,pc of][]{1931AnLun...2....1C} and an age of 2 to 3\,Gyr, from which they concluded that the cluster is real but sparse because it is dissolving into the Galactic field. 

Later the same year, \citet{2000A&A...357..145C} argue the opposite: based on $BVI$ photometry, NGC~6994 is not a physical object but a chance alignment of a handful of bright stars on the same line of sight. 
\citet{2001A&A...366..827B} still included it in a list of possible open cluster remnants. \citet{2002A&A...383..163O} had what might have seemed like the final word on the matter by showing the six brightest proposed members do not shared a common proper motion (with Tycho~2 data) or radial velocity, and that these stars are therefore not related. The compiled catalogues of \citet{2002A&A...389..871D} and \citet{2013A&A...558A..53K} do not include it. \citet{2006BASI...34..153C} mention NGC~6994 as an object whose story `represents a real lesson'. \citet{2007A&A...468..139P} include it in their list of open cluster remnants but surmise that it is probably a field fluctuation.

The non-physicality of NGC~6994 was recently confirmed again by \citet{2018MNRAS.480.5242K} and \citet{2018A&A...618A..93C} on the basis of additional radial velocities and \textit{Gaia}~DR2 astrometry and no doubt should remain that M~73/NGC~6994 is an asterism.

\subsection{NGC~7036}
This object was marked non-existent by \citet{1973rncn.book.....S} and is not present in the WEBDA database. \citet{2002A&A...385..471C} also expressed doubts about the reality of this cluster.
The catalogues of DAML and MWSC quote distances of 1000 and 1036\,pc. \citet{2006A&A...446..949D} and \citet{2010A&A...516A...3K} find potential members in proper motion space, but do not conclude on the reality of the object.
\citet{2001A&A...366..827B} and \citet{2019AJ....157...12B} consider it an open cluster remnant.

\begin{figure}[ht]
\begin{center} \resizebox{0.49\textwidth}{!}{\includegraphics[scale=0.6]{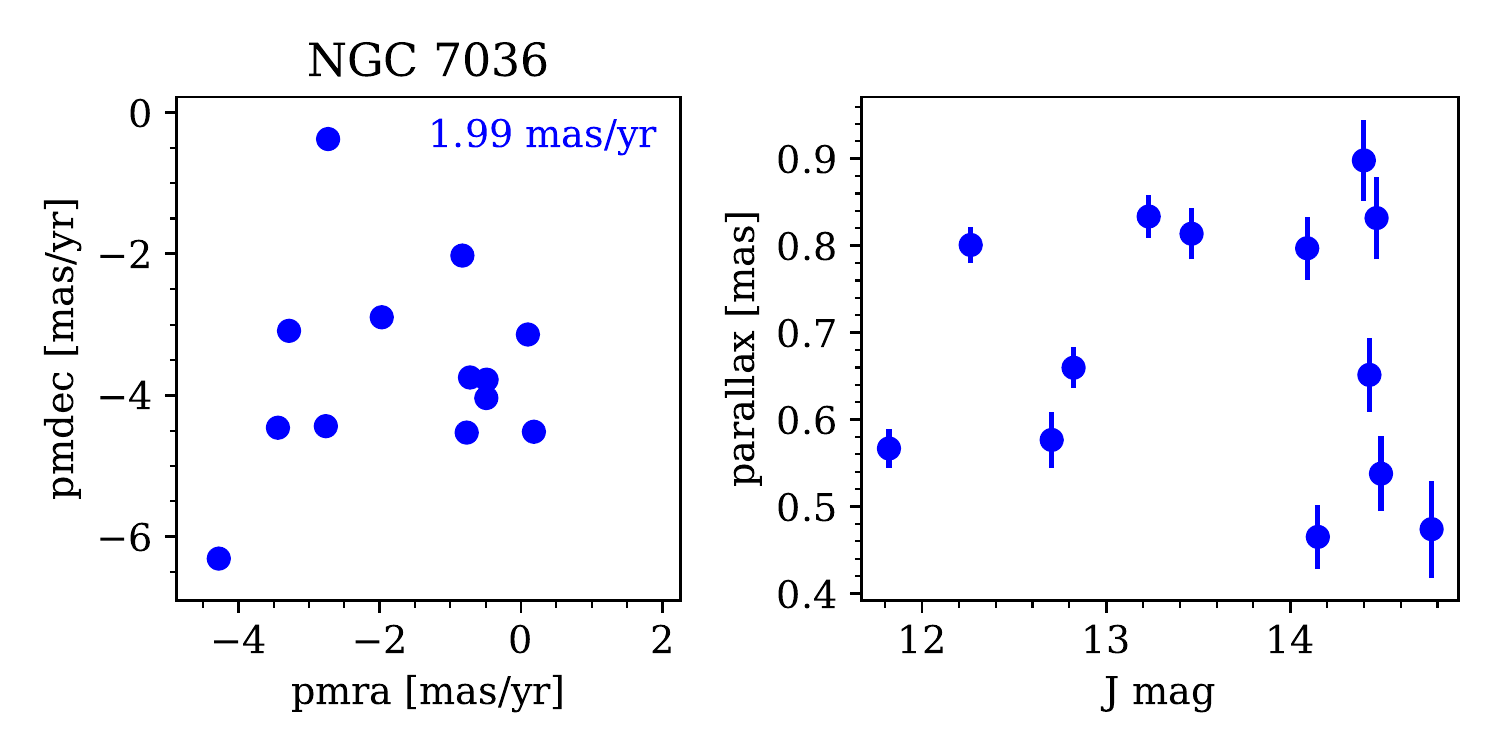}} \caption{\label{fig:angelo19members_NGC_7036} NGC~7036. Left: \textit{Gaia}~DR2 proper motions for the members proposed by \citet{2019A&A...624A...8A}. The error bars are smaller than the markers. The total proper-motion dispersion is indicated. Right: parallax vs 2MASS $J$ mag for the same stars. } \end{center}
\end{figure}

This object was also studied by \citet{2019A&A...624A...8A}, who propose 13 possible members. Figure~\ref{fig:angelo19members_NGC_7036} shows that these stars do not form a coherent group in either proper motion or parallax space.

\subsection{NGC~7055}
This object was marked non-existent by \citet{1973rncn.book.....S} and is not present in the WEBDA database nor in the MWSC catalogue. The DAML catalogue quotes a distance of 1275\,pc and age of 800\,Myr \citep[after][]{2011JKAS...44....1T}, while \citet{2012A&A...542A..68P} estimate an age of $\sim$100\,Myr and a distance of 3300\,pc.

\subsection{NGC~7084} 
Reported by J.~Herschel as a `coarse scattered cluster', this object was marked non-existent by \citet{1973rncn.book.....S}, and is not present in the WEBDA database.
The catalogue of DAML quote a distance of 765\,pc and $\log t$=9.18 \citep[after][]{2011JKAS...44....1T}, while MWSC quotes 1259\,pc and $\log t$=9.425. 

\citet{2010A&A...516A...3K} report that their procedure intended to provide a mean proper motion for this cluster returned a poor fit, which confirms the non-existence of this cluster. 

\citet{2019AJ....157...12B} flagged this object as an open cluster remnant.

\subsection{NGC~7127}
This object is not listed in the WEBDA database, nor in the MWSC catalogue. The DAML catalogue quotes a distance of 1445\,pc and age of 400\,Myr \citep[after][]{2011JKAS...44....1T} while \citet{2012A&A...542A..68P} estimate a much younger age of $\sim$10\,Myr, and a much larger distance of 5700\,pc.

\subsection{NGC~7193}

This object was marked non-existent by \citet{1973rncn.book.....S} and is not present in the WEBDA database nor in the MWSC catalogue.

\begin{figure}[ht]
\begin{center} \resizebox{0.49\textwidth}{!}{\includegraphics[scale=0.6]{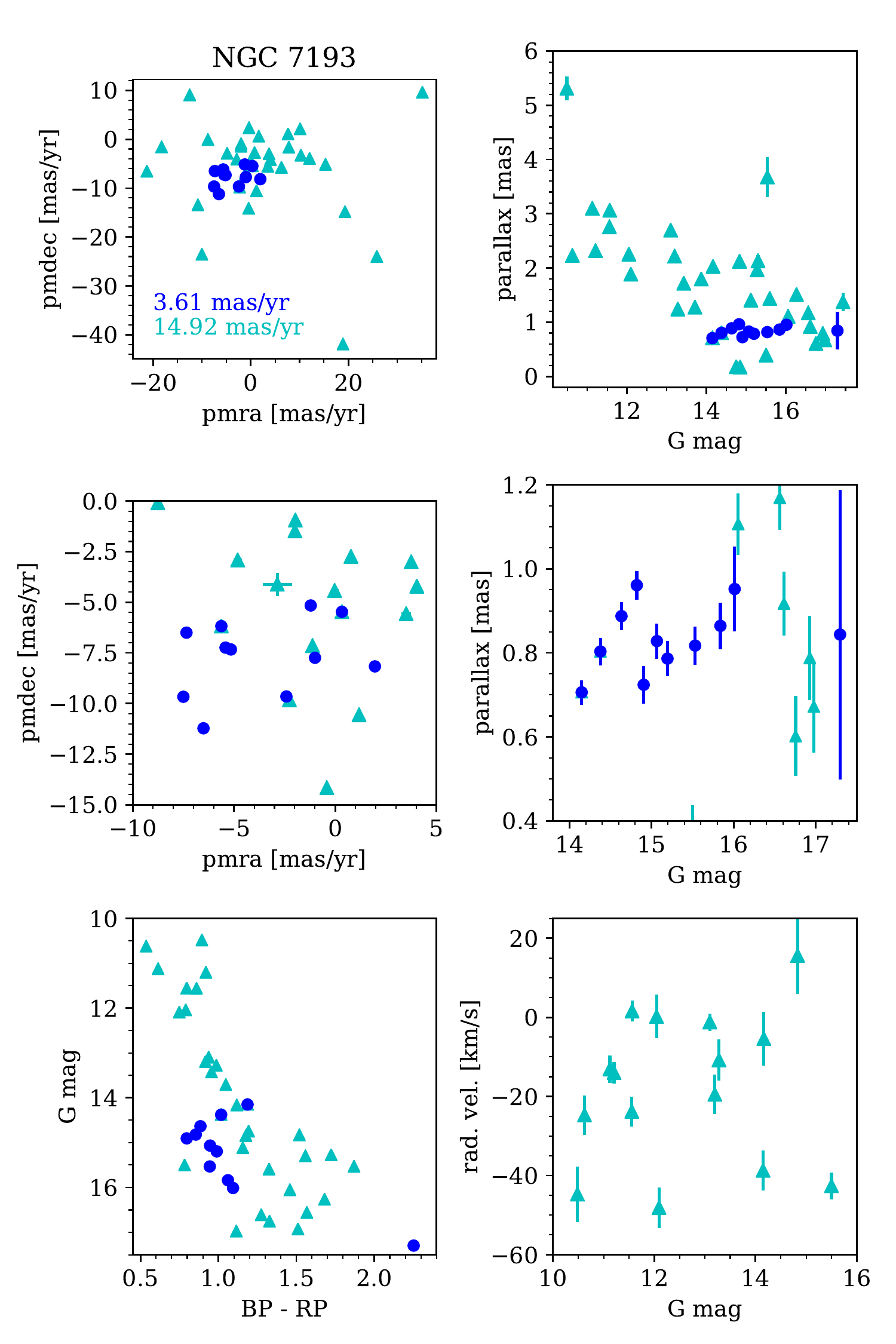}} \caption{\label{fig:angelo19members_NGC_7193} NGC~7193. Top left: \textit{Gaia}~DR2 proper motions for the members proposed by \citet{2019A&A...624A...8A} (blue dots) and \citet{2017RAA....17....4D} (cyan triangles). The error bars are smaller than the markers. The total proper-motion dispersion is indicated for both samples. Top right: parallax vs $G$ mag for the same stars. Middle row: same as top row, with a different scale. Bottom left: \textit{Gaia}~DR2 CMD for the same stars. Bottom right: radial velocity vs $G$ magnitude for the \citet{2017RAA....17....4D} stars.} \end{center}
\end{figure}

\citet{2011JKAS...44....1T} estimate an age of 4.5\,Gyr and a distance of 1080\,pc, while \citet{2017RAA....17....4D} find a distance of 501\,pc. Such a nearby object would be very difficult to miss with \textit{Gaia} data. Its location near $(\ell,b)=(70.1,-34.3)$ would also make it virtually unaffected by interstellar extinction and easily visible as a tight sequence in a CMD. The stars which \citet{2017RAA....17....4D} consider probable members of the cluster exhibit a very large radial velocity dispersion (Fig.~\ref{fig:angelo19members_NGC_7193}), interpreted by the authors as a consequence of the increasing binary fraction as clusters evolve. However, the \textit{Gaia} proper motions are relatively unaffected by unresolved binaries \citep{2018A&A...616A..17A} and should therefore exhibit a compact distribution if the group truly was a cluster remnant.

Cross-matching the \citet{2017RAA....17....4D} members with the \textit{Gaia}~DR2 catalogue reveals no trace of a coherent group in proper motion space (Fig.~\ref{fig:angelo19members_NGC_7193}). These stars are coincidentally aligned on the same line of sight but they have parallaxes ranging from 0.1 to over 3\,mas (with typical uncertainties of 0.05\,mas). 

\citet{2019A&A...624A...8A} report 11 members from \textit{Gaia}~DR2, with only two in common with the 34 candidates of \citet{2017RAA....17....4D}. Figure~\ref{fig:angelo19members_NGC_7193} shows they do not form a coherent cluster in \textit{Gaia}~DR2 astrometry either. In the same figure, we also see that although the members of \citet{2017RAA....17....4D} might form a contaminated but plausible cluster sequence in a CMD, those of \citet{2019A&A...624A...8A} clearly do not.

\subsection{NGC~7772}

NGC~7772 appears as a compact aggregate of six stars of roughly equal magnitude, and a seventh brighter and redder star.
\citet{1971A&A....13..309W} expressed doubts on the reality of this sparse aggregate as a true cluster, that \citet{2001A&A...366..827B} classified as an open cluster remnant.
The first deep investigation of NGC~7772 was the study of \citet{2002A&A...385..471C}, who identified possible members and also classified it as a cluster remnant, with an age of $\sim$1.5\,Gyr. \citet{2013A&A...558A..53K} report an age of over 1\,Gyr and a distance of 1250\,pc. Located at $b$$\sim$-44$^{\circ}$, this object would be an old, high-altitude cluster.
 
\citet{2010A&A...516A...3K} remarked that using the PM2000 Bordeaux proper motion catalogue \citep[with precisions from 1.5 to 6\,mas\,yr$^{-1}$,][]{2006A&A...448.1235D}, only two stars appear to have consistent proper motions within their nominal uncertainties, while another two (located near them on the sky) might be members if the system had a large intrinsic proper-motion dispersion of several mas\,yr$^{-1}$.

\citet{2018MNRAS.480.5242K}, using \textit{Gaia}~DR2 data supplemented with radial velocities from high-resolution spectroscopy, remark that not a single pair of stars among the proposed members of \citet{2002A&A...385..471C} and \citet{2013A&A...558A..53K} have matching parameters. \citet{2019A&A...624A...8A} flag it as an asterism.

\subsection{NGC~7801}

This object is not listed in the WEBDA database and was marked as non-existent by \citet{1973rncn.book.....S}. 
\citet{2004yCat.7239....0C} notes that this object has always been listed as either an asterism or with a question mark. \citet{2011JKAS...44....1T} estimate an age of 1.7\,Gyr and a distance of $\sim$1400\,pc. NGC~7801 is listed as a remnant by DAML, and as a 2\,Gyr open cluster at a distance of 1953\,pc in MWSC.
\citet{2018MNRAS.473..849D} included it in their study of mass-segregated clusters.

\subsection{NGC~7826}

This object, originally described as `a cluster of a few coarsely scattered large stars' by W. Herschel, was classified as non-existent by \citet{1973rncn.book.....S} and it is not present in the catalogue of \citet{2013A&A...558A..53K} or in the WEBDA database. It is however listed in DAML with an age of 2\,Gyr and a distance of 620\,pc \citep[after][]{2011JKAS...44....1T}.

\citet{2018MNRAS.480.5242K}, supplementing \textit{Gaia}~DR2 data with radial velocities from high-resolution spectroscopy, conclude that this object is not a physical clusters of related stars.

\subsection{IC~1023} 
This object is not included in the WEBDA database and is absent from the DAML catalogue, but is listed in MWSC at a distance of 1298\,pc ($\log t$=9.48). \citet{2001A&A...366..827B} and  \citet{2019AJ....157...12B} consider it as an open cluster remnant.

\subsection{Ruprecht~3}

\begin{figure*}[ht]
\begin{center} \resizebox{0.99\textwidth}{!}{\includegraphics[scale=0.6]{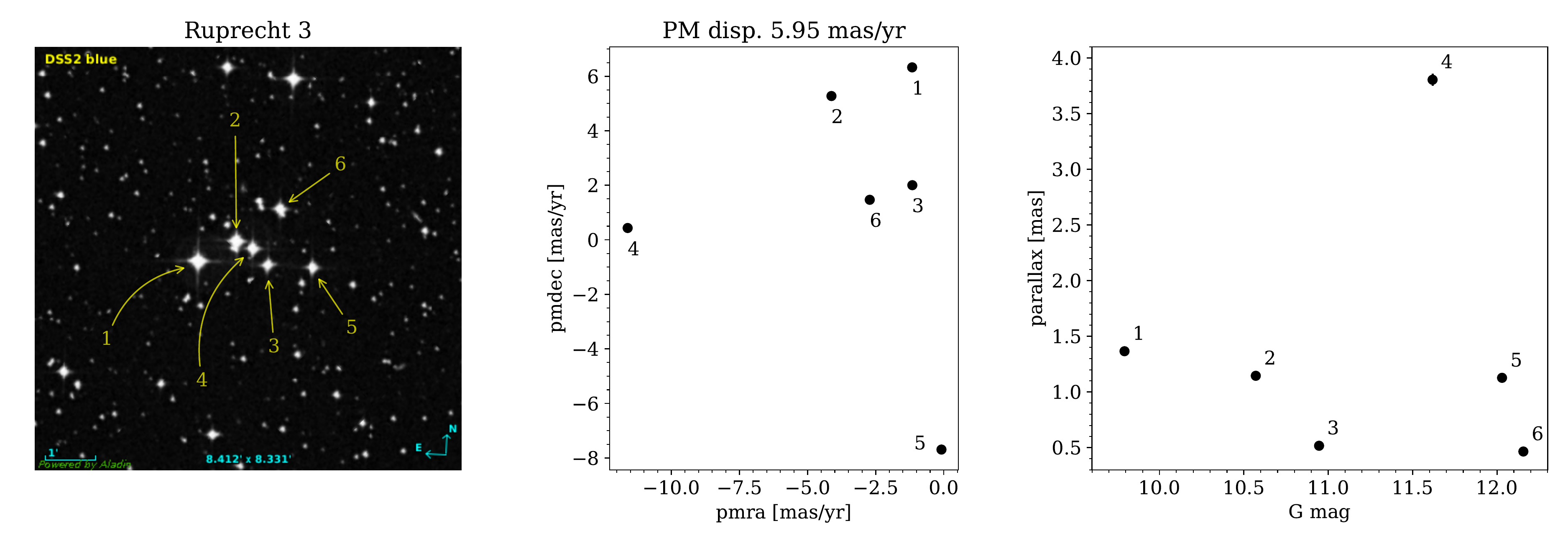}} \caption{\label{fig:rup3_bright} Left: DSS2 image of the asterism Ruprecht~3, indicating the six brightest stars. Middle: \textit{Gaia}~DR2 proper motions for these stars. Right: \textit{Gaia}~DR2 parallax and $G$ magnitude for the same stars. Proper motion and parallax error bars are smaller than the symbols.} \end{center}
\end{figure*}

Visually, this asterism appears as a tight concentration of six bright stars (see Fig.~\ref{fig:rup3_bright}). The object is present in WEBDA, but with no listed parameters. The MWSC catalogue cites a distance of 1259\,pc and $\log t$=9.1 and DAML cites comparable parameters of 1100\,pc and $\log t$=9.18.

\citet{2003A&A...399..113P} estimate a distance of 685 to 760\,pc. \citet{2004A&A...427..485B} compare Ruprecht~3 to NGC~1252, NGC~7036, and NGC~7772, all three of which were later shown to be non-physical objects. Ruprecht~3 is present in the list of old clusters studied by \citet{2007A&A...468..139P}. 

\citet{2017MNRAS.466..392P} noticed that the \textit{Gaia}~DR1 parallaxes of the five brightest potential members of Ruprecht~3 differ too much for them being members of the same cluster, and conclude that the object must be an asterism.

\begin{figure}[ht]
\begin{center} \resizebox{0.49\textwidth}{!}{\includegraphics[scale=0.6]{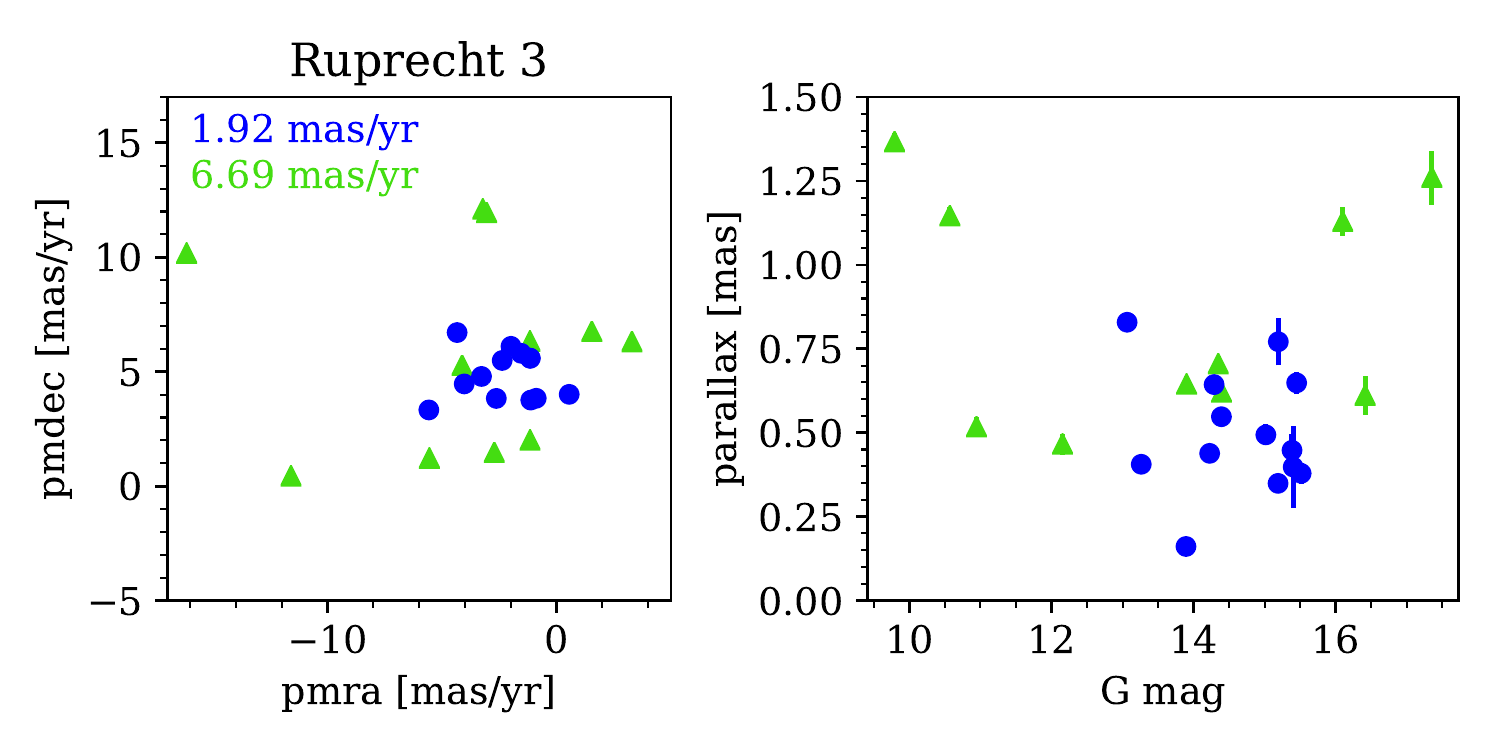}} \caption{\label{fig:angelo19members_Ruprecht_3} Ruprecht~3: Left: \textit{Gaia}~DR2 proper motions for the members proposed by \citet{2019A&A...624A...8A} (blue dots) and \citet{2003A&A...399..113P} (green triangles). The error bars are smaller than the markers. The total proper-motion dispersion is indicated for both samples. Right: parallax vs $G$ mag for the same stars. } \end{center}
\end{figure}

\citet{2003A&A...399..113P} propose 11 members of this object, while \citet{2019A&A...624A...8A} report 14 members. These two studies have no members in common. Figure~\ref{fig:angelo19members_Ruprecht_3} shows that none of the proposed members form a coherent cluster in \textit{Gaia}~DR2 astrometry.

\subsection{Ruprecht~46}

Ruprecht~46 is not listed in DAML but it is included in MWSC, with an estimated distance of 1467\,pc. \citet{1995MNRAS.276..563C} pointed out that the region does correspond to a density enhancement of stars brighter than $V$$\sim$14.5, but presents no meaningful feature in a CMD.

\subsection{Ruprecht~155}
This object has an entry in the WEBDA database, but no associated parameters. It is absent from the DAML catalogue but listed in MWSC, at a distance of 2300\,pc for a $\log t$ of 8.5. 
\citet{2019AJ....157...12B} flagged it as an open cluster remnant. This object was never the target of a dedicated study, and is erroneously mentioned in the abstract of \citet{2018MNRAS.475.4122S}, who studied Ruprecht~175.

\subsection{Collinder~471}
This object has an entry in the WEBDA database, but no associated parameters. 
The catalogues of DAML and MWSC quote distances of 2003 and 2210\,pc, and $\log t$=6.88 and 8.8 (respectively). \citet{2018MNRAS.475.4122S} remark that these two catalogues also list very discrepant apparent radii of 65\,arcmin and 8.4\,arcmin (respectively) and that themselves are unable to determine a radius that makes this stellar group apparent in proper motion. \citet{2019AJ....157...12B} flag it as an embedded cluster.

\subsection{Basel~5}

The oldest available study of Basel~5 seems to be that of \citet{1966ZA.....64...67S}, who estimate a distance of 850\,pc but remark that the aggregate is `somewhat irregular in shape' and `rather elongated'. This object is listed in DAML at a distance of 766\,pc and in MWSC at a distance of 995\,pc.
\citet{2019MNRAS.488.1635A} include Basel~5 in their study of evolved clusters with \textit{Gaia}~DR2. The proper motion and parallax diagram they show for this object are much more dispersed than what one would expect from a physical object. They estimate a distance of 1.74\,kpc from a CMD but fail to remark that the mean parallax of those stars is $\sim$0.3\,mas, which would indicate a distance at least twice as large. 

\begin{figure*}[!htb]
   \begin{minipage}{0.48\textwidth}
     \centering
     \includegraphics[width=.9\linewidth]{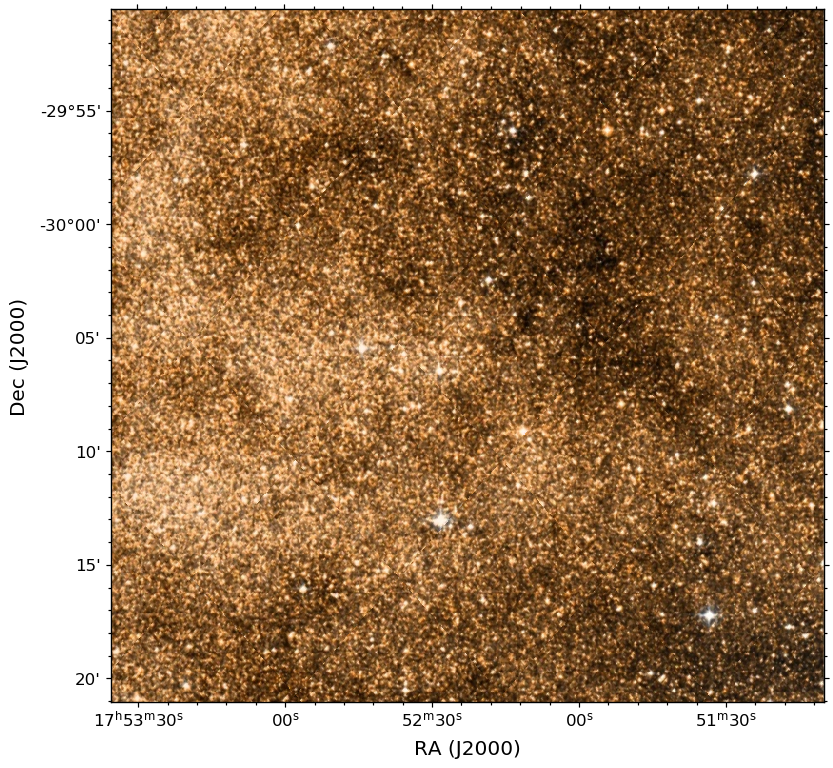}
   \end{minipage}\hfill
   \begin{minipage}{0.48\textwidth}
     \centering
     \includegraphics[width=.9\linewidth]{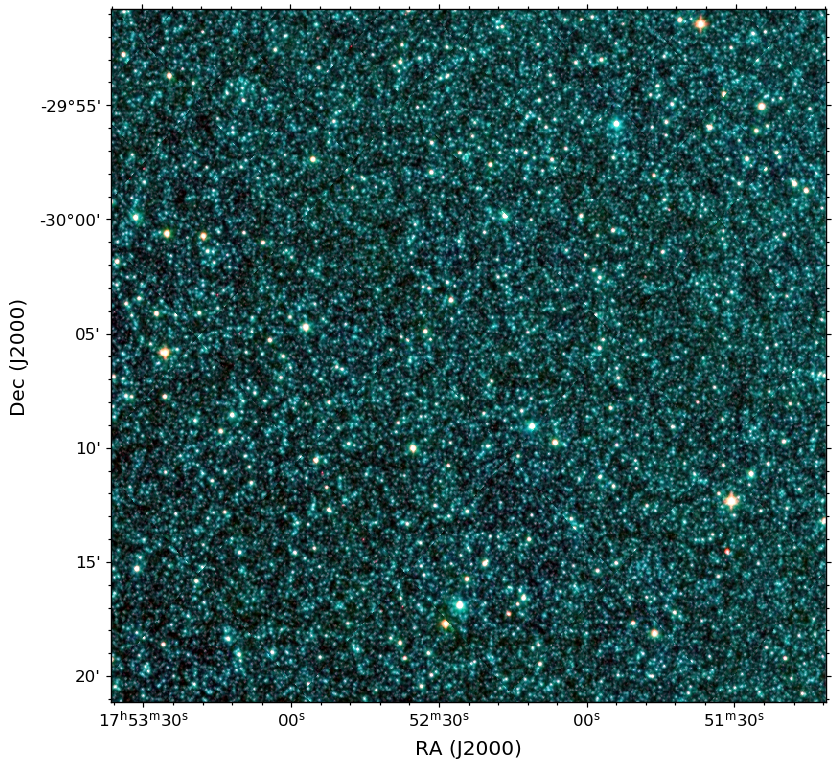}
   \end{minipage}

\caption{\label{fig:basel5} Images from DSS2 (left) and Spitzer/GLIMPSE \citep[][right]{2009PASP..121..213C} for the same field of about 25$\times$25\,arcmin centred on the asterism Basel~5.}

\end{figure*}

Given its Galactic coordinates of $(\ell,b)=(359.8^{\circ},-1.9^{\circ})$, this group is projected against a very dense background shaped by patchy extinction (Fig.~\ref{fig:basel5}) and is likely to be an asterism caused by extinction patterns.

\subsection{Loden~1}  \label{indiv:end}

\citet{1980A&AS...41..173L} originally reported this grouping of stars as a candidate cluster. \citet{2005A&A...438.1163K} estimated an age of 2\,Gyr and a distance of 360\,pc for this object, making it one of the oldest nearby clusters. These numbers were revised to 786\,pc and an age of 200\,Myr in the MWSC catalogue.

Remarking that the cluster was not visible in a CMD, \citet{2016AJ....152....7H} collected radial velocities to better identify its members. They finally determined that no co-moving group of old stars is present in the direction of Loden~1.

\section{Six representative UPK clusters}

\begin{figure*}[ht]
\begin{center} \resizebox{0.9\textwidth}{!}{\includegraphics[scale=0.6]{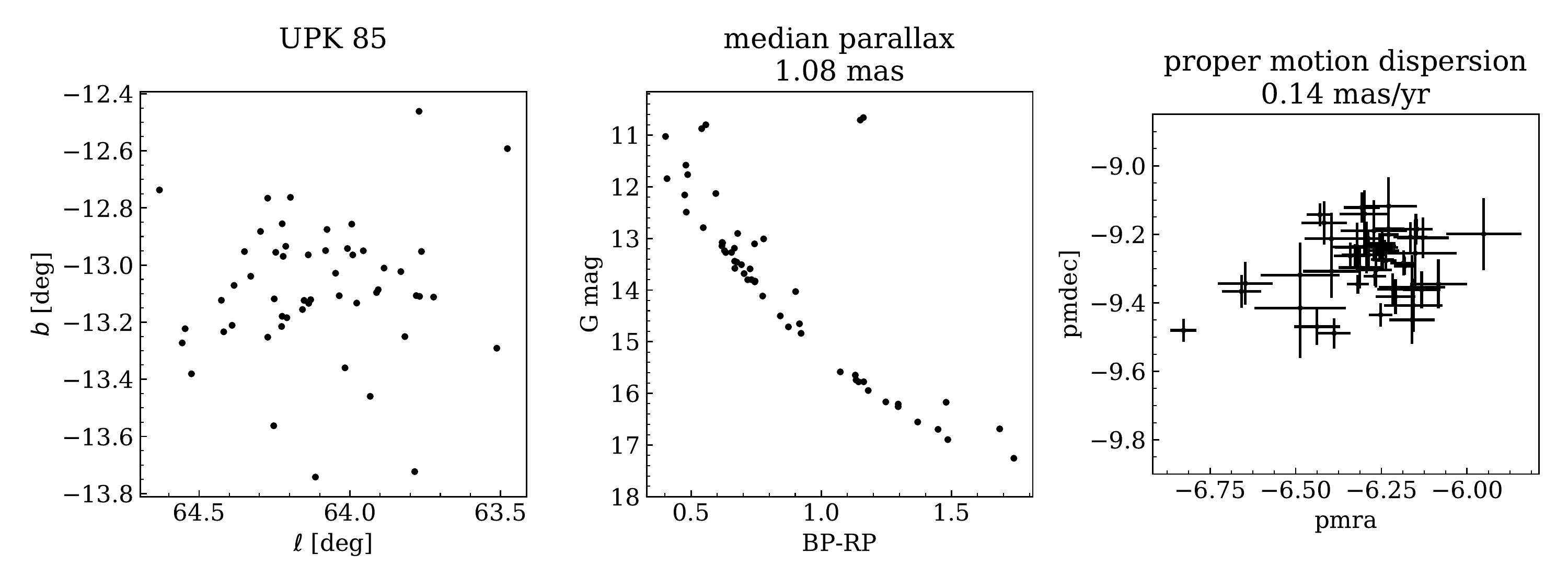}} \caption{\label{fig:upk_85} Probable members (probability > 70\%) of UPK~85. Left: sky position. Middle: colour-magnitude diagram. Right: proper motions.} \end{center}
\end{figure*}

\begin{figure*}[ht]
\begin{center} \resizebox{0.9\textwidth}{!}{\includegraphics[scale=0.6]{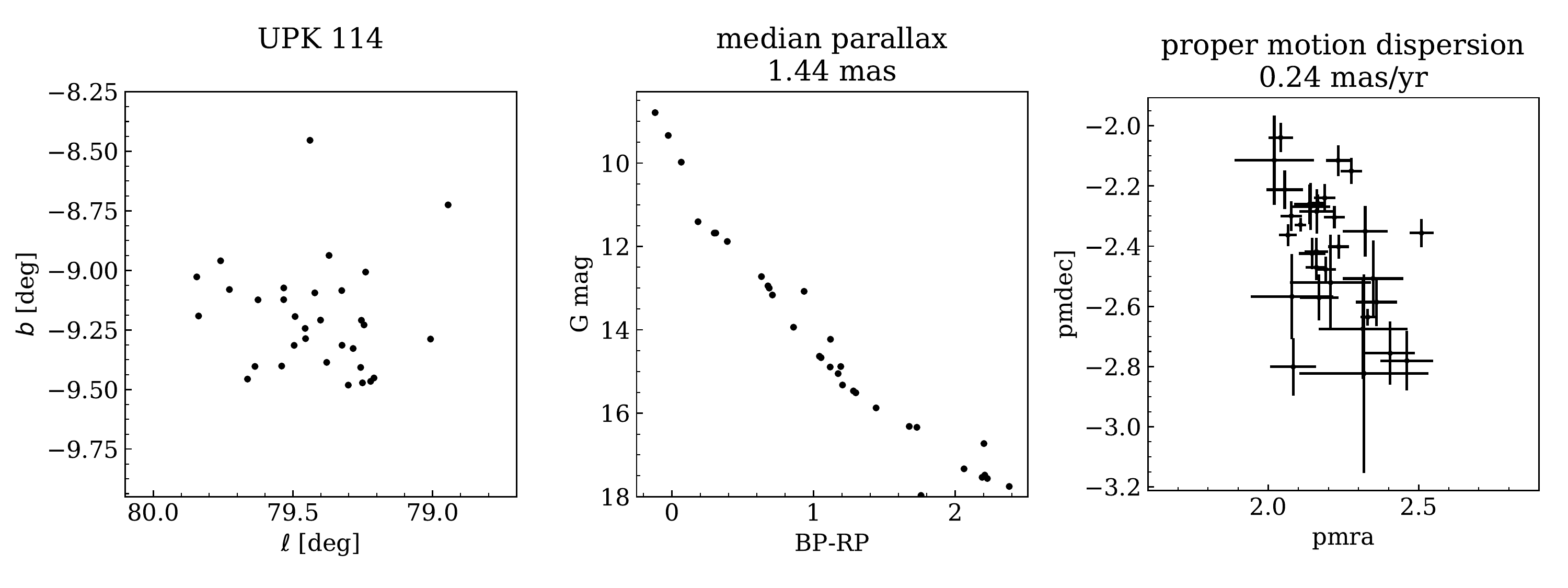}} \caption{\label{fig:upk_144} Same as Fig.~\ref{fig:upk_85}, for UPK~114.} \end{center}
\end{figure*}

\begin{figure*}[ht]
\begin{center} \resizebox{0.9\textwidth}{!}{\includegraphics[scale=0.6]{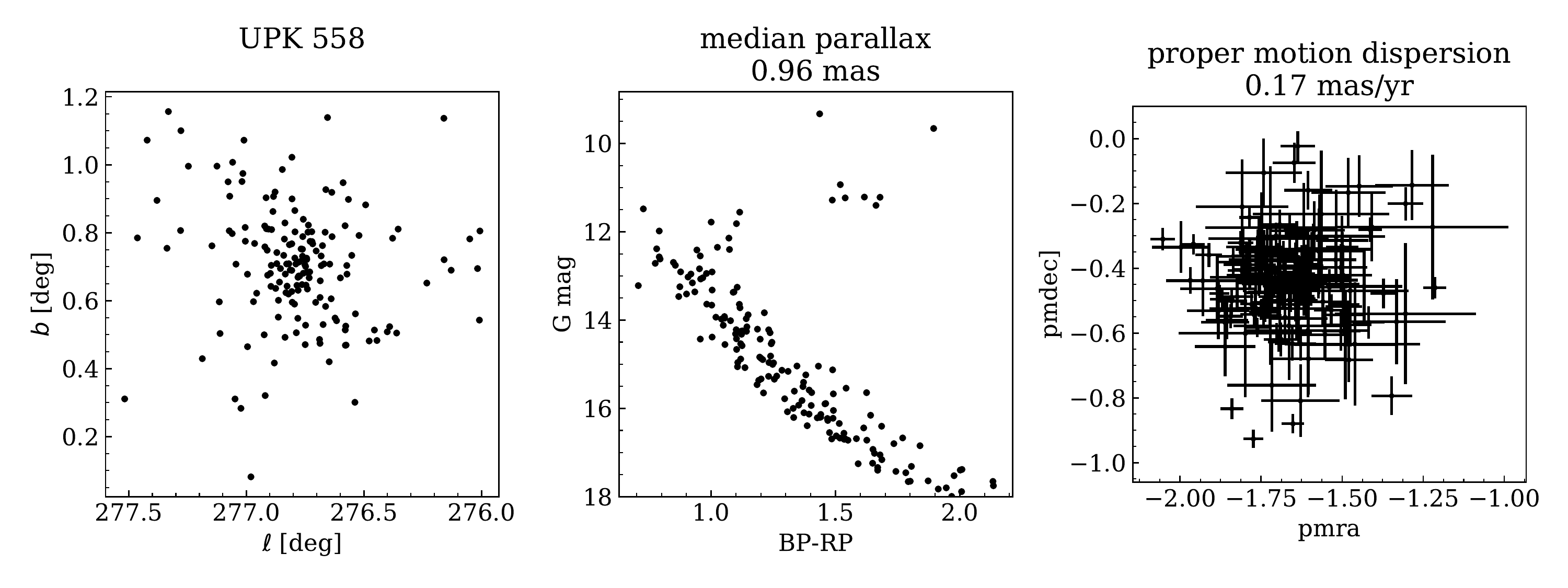}} \caption{\label{fig:upk_558} Same as Fig.~\ref{fig:upk_85}, for UPK~558.} \end{center}
\end{figure*}

\begin{figure*}[ht]
\begin{center} \resizebox{0.9\textwidth}{!}{\includegraphics[scale=0.6]{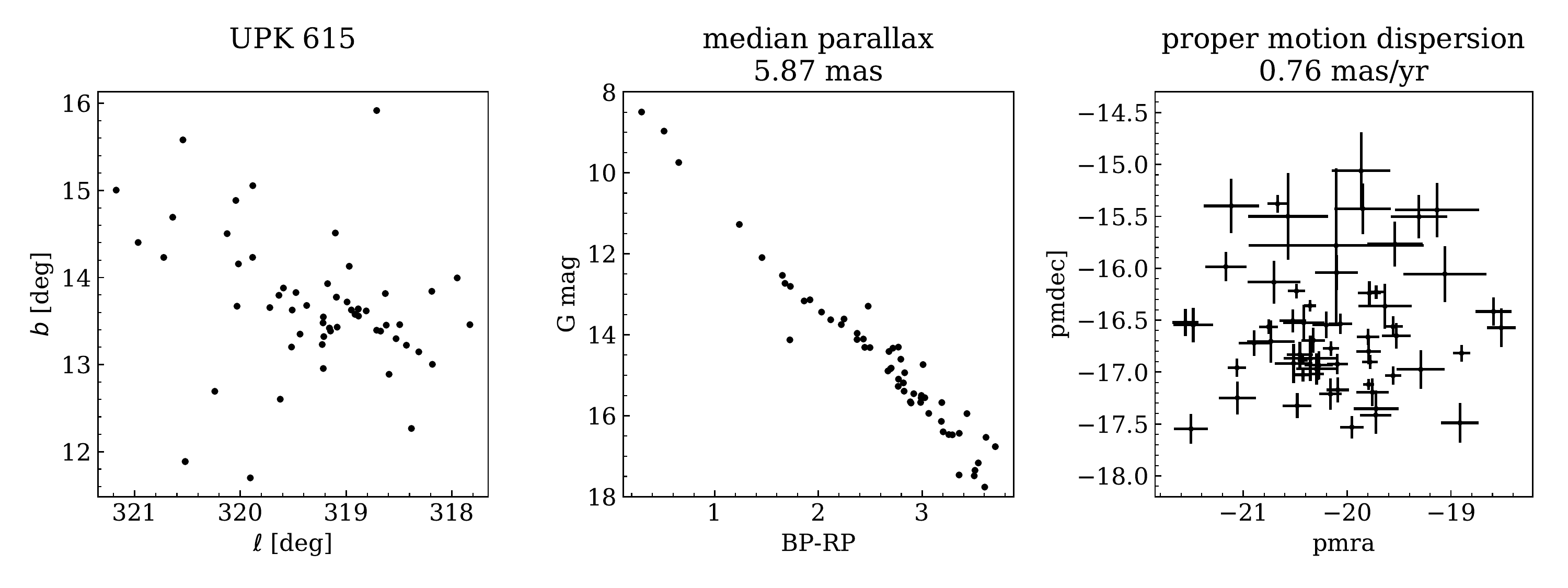}} \caption{\label{fig:upk_615} Same as Fig.~\ref{fig:upk_85}, for UPK~615.} \end{center}
\end{figure*}

\begin{figure*}[ht]
\begin{center} \resizebox{0.9\textwidth}{!}{\includegraphics[scale=0.6]{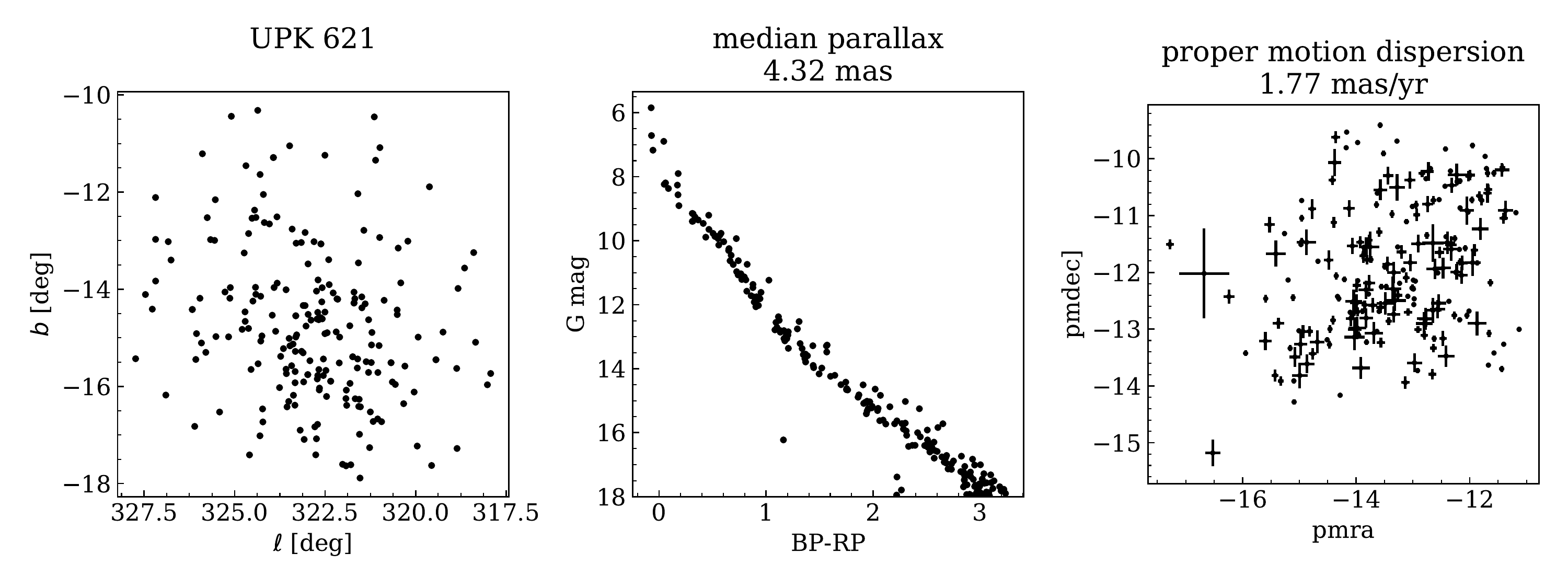}} \caption{\label{fig:upk_621} Same as Fig.~\ref{fig:upk_85}, for UPK~621.} \end{center}
\end{figure*}

\begin{figure*}[ht]
\begin{center} \resizebox{0.9\textwidth}{!}{\includegraphics[scale=0.6]{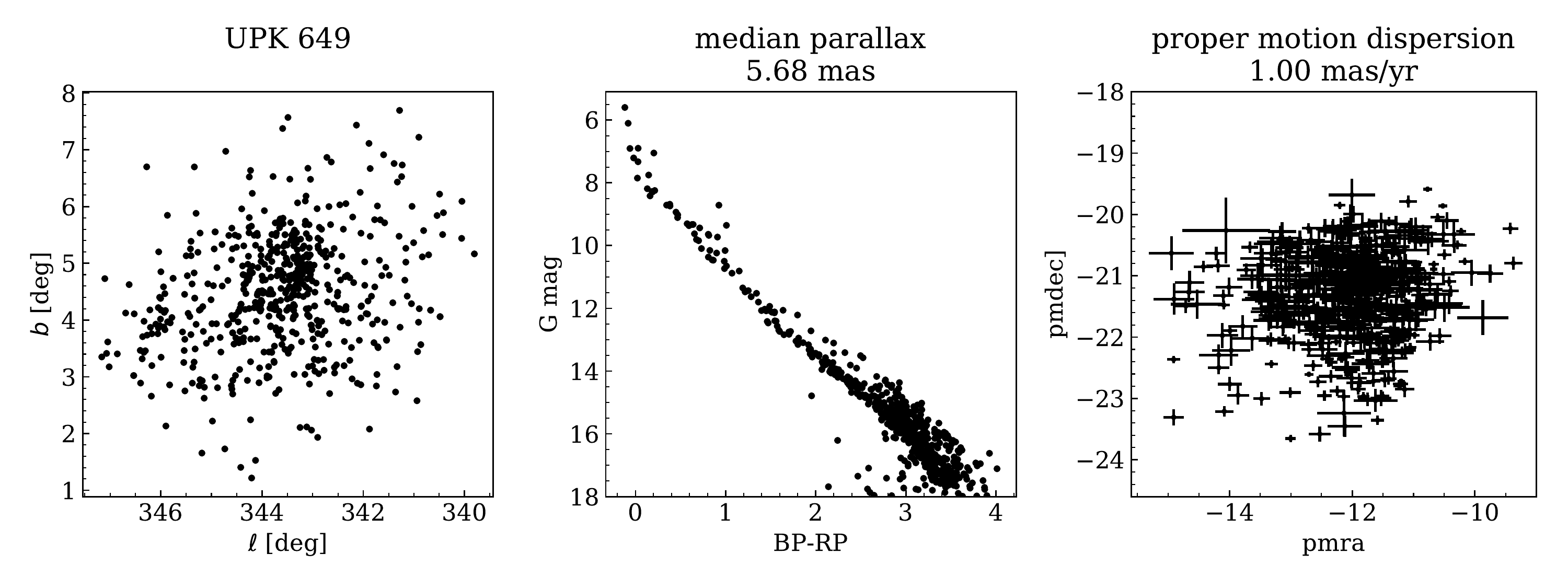}} \caption{\label{fig:upk_649} Same as Fig.~\ref{fig:upk_85}, for UPK~649.} \end{center}
\end{figure*}

\section{Decontaminating a CMD with an offset field}
We mention in Sect.~\ref{sec:signal} that comparing the colour-magnitude diagram (CMD) of an assumed cluster with a surrounding reference field can `reveal' an apparent cluster sequence even when there is no cluster at all. Such artificial detections are especially likely in the presence of variable extinction. \citet{2007MNRAS.377.1301B} warn that when working with a binned CMD, any differential reddening whose effect is comparable to the bin dimension will lead to erroneous comparisons between the assumed cluster and reference fields.

\begin{figure*}[ht]
\begin{center} \resizebox{0.95\textwidth}{!}{\includegraphics[scale=0.6]{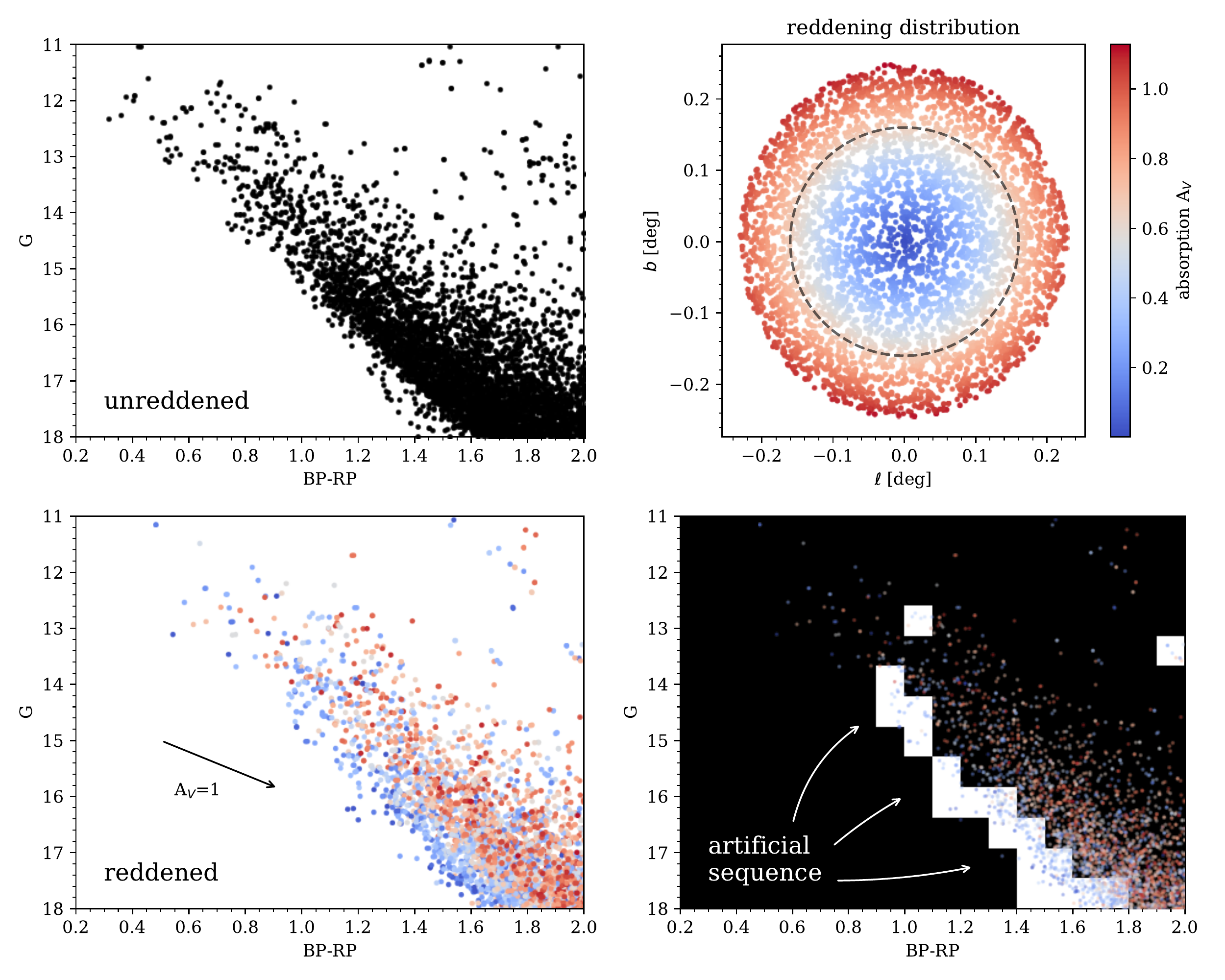}} \caption{\label{fig:decontamination} Top left: CMD of a simulated field population. Top right: spatial distribution of the reddening added the simulated field. The dashed line indicates the inner region. Bottom left: CMD after adding reddening. The arrows indicates A$_V$=1. Bottom right: the bins marked in white correspond to areas where the inner region CMD is denser than the outer region CMD by more than two sigma (assuming Poisson noise).} \end{center}
\end{figure*}

Figure~\ref{fig:decontamination} shows the CMD of a synthetic field population obtained from the \textit{Gaia} Universe Snapshot Model \citep[GUMS,][]{2012A&A...543A.100R}. In this simple experiment we assign reddening\footnote{For convenience, we approximate A$_G$=0.84\,A$_V$ and E$_{BP-RP}$=0.42\,A$_V$.} to stars as a function of their distance from the centre of the field of view. We divide the reddened CMDs of the inner and outer region (chosen to be of equal area) into bins of 0.5\,mag in $G$ magnitude and 0.1 in $BP$-$RP$ colour. The bottom right panel of Fig.~\ref{fig:decontamination} highlights the bins where the inner-region CMD is denser than the outer-region CMD by more than two sigma (assuming Poissonian uncertainties on the counts in each bin). They align in a sequence following the blue edge of the inner-region CMD.

\end{document}